\documentclass[journal,10pt]{IEEEtran}

\usepackage[hyphens]{url}
\usepackage{hyperref}
\hypersetup{breaklinks=true}

\usepackage{amsmath}
\usepackage{nccmath}
\usepackage[utf8]{inputenc} 
\usepackage{epsfig,scalefnt,multirow}
\usepackage{url}
\usepackage{ifthen}
\usepackage{mathtools}
\usepackage{cite}
\usepackage{graphicx}
\usepackage{amssymb}
\usepackage{tabularx}
\usepackage{amsmath}
\usepackage{epstopdf}
\usepackage{epsf}
\usepackage{algorithm}
\usepackage{algpseudocode}
\usepackage{algpascal}
\usepackage{cases}
\usepackage{caption}
\usepackage[labelformat=simple]{subcaption}

\usepackage{stfloats}
\usepackage{float}
\usepackage{xcolor}
\usepackage{tabularx}

\usepackage{epsfig,amsmath,amssymb,epsf,amsthm,scalefnt,multirow}
\usepackage{xcolor}
\usepackage{float}
\usepackage{cite}
\usepackage{psfrag}
\usepackage{gensymb}
\usepackage{cleveref}
\usepackage{mathrsfs}
\usepackage{mathtools}
\usepackage{algpseudocode}
\usepackage{algorithm}
\usepackage{enumerate}

\newsavebox{\imagebox}

\newtheorem*{lemma*}{Lemma}

\def\diag{\mathop{\mathrm{diag}}}

\def\b0{{\pmb{0}}} 

\def\ba{{\mathbf{a}}} \def\bb{{\mathbf{b}}}  
  \def\bg{{\mathbf{g}}} 
   
 \def\bn{{\mathbf{n}}}  \def\bp{{\mathbf{p}}}
\def\bq{{\mathbf{q}}}   \def\bt{{\mathbf{t}}}
   \def\bx{{\mathbf{x}}}
\def\by{{\mathbf{y}}}   

\def\bA{{\mathbf{A}}} \def\bB{{\mathbf{B}}}  
 \def\bF{{\mathbf{F}}} \def\bG{{\mathbf{G}}} \def\bH{{\mathbf{H}}}
\def\bI{{\mathbf{I}}} \def\bJ{{\mathbf{J}}}  \def\bL{{\mathbf{L}}}
\def\bM{{\mathbf{M}}} \def\bN{{\mathbf{N}}}  
   \def\bT{{\mathbf{T}}}

\begin{document}

\title{Multi-User SLNR-Based Precoding With Gold Nanoparticles in Vehicular VLC Systems}

\author{Geonho Han,~\IEEEmembership{Member,~IEEE}, Hyuckjin Choi,~\IEEEmembership{Member,~IEEE}, Hyesang Cho,~\IEEEmembership{Member,~IEEE},\\ Jeong Hyeon Han, Ki Tae Nam, and Junil Choi,~\IEEEmembership{Senior Member,~IEEE}
	\thanks{This work was partly supported by Agency for Defense Development (No.912909601, Challengeable Future Defense Technology Research and Development Program in 2022, 50\%) and Institute of Information \& communications Technology Planning \& Evaluation (IITP) grant funded by the Korea government (MSIT) (No.RS-2024-00444230, Development of Wireless Technology for Integrated Sensing and Communication, 50\%).
		
		G. Han is with the 6G Wireless Access System Research Division, Electronics and Telecommunications Research Institute (ETRI), Daejeon 34129, South Korea (e-mail: ghhan6@etri.re.kr).
        
        H. Choi, H. Cho, and J. Choi are with the School of Electrical Engineering, Korea Advanced Institute of Science and Technology, Daejeon 34141, South Korea (e-mail: \{hugzin008, nanjohn96, junil\}@kaist.ac.kr).
		
		J. H. Han and K. T. Nam are with the Department of Materials Science and Engineering, Seoul National University, Seoul 08826, South Korea (e-mail: hanjh@snu.ac.kr, nkitae@snu.ac.kr).
}}

{}

\maketitle

\begin{abstract}
Visible spectrum is an emerging frontier in wireless communications for enhancing connectivity and safety in vehicular environments. The vehicular visible light communication (VVLC) system is a key feature in leveraging existing infrastructures, but it still has several critical challenges. Especially, VVLC channels are highly correlated due to the small gap between light emitting diodes (LEDs) in each headlight, making it difficult to increase data rates by spatial multiplexing. In this paper, we exploit recently synthesized gold nanoparticles (GNPs) to reduce the correlation between LEDs, i.e., the chiroptical properties of GNPs for differential absorption depending on the azimuth angle of incident light are used to mitigate the LED correlation. In addition, we adopt a signal-to-leakage-plus-noise ratio (SLNR)-based precoder to support multiple users. The ratio of RGB light sources in each LED also needs to be optimized to maximize the sum SLNR satisfying a white light constraint for illumination since the GNPs can vary the color of transmitted light by the differential absorption across wavelength. The nonconvex optimization problems for precoders and RGB ratios can be solved by the generalized Rayleigh quotient with the approximated shot noise and successive convex approximation (SCA). The simulation results show that the SLNR-based precoder with the optimized RGB ratios significantly improves the sum rate in a multi-user vehicular environment and the secrecy rate in a wiretapping scenario. The proposed SLNR-based precoding verifies that the decorrelation between LEDs and the RGB ratio optimization are essential to enhance the VVLC performance.
\end{abstract}\

\renewcommand\IEEEkeywordsname{Index Terms}

\section{Introduction}\label{sec_intro}
Vehicular communications become essential to enhance safety, improve traffic efficiency, and enable autonomous driving in next generation wireless communications \cite{Choi:2016}. It is typical to employ radio frequency (RF) spectra in vehicular communications while there are several limitations such as the shortage of available spectrum, and severe interference and channel congestion in high density traffic scenarios. Visible spectrum can be a complementary solution thanks to i) low interference by high directivity and ii) less implementation cost by exploiting existing infrastructures \cite{Cuailean:2017,Memedi:2020}. Especially, due to the widespread adoption of light emitting diodes (LEDs) equipped in headlights, taillights, and streetlamps for illumination, vehicular visible light communication (VVLC) systems are recently being studied to establish the intelligent transportation system (ITS) as overcoming the limited communication range and outdoor conditions including high mobility, fluctuating weather, and dynamic traffic density \cite{Karbalayghareh:2020,Alsalami:2021,Luo:2015}.

\subsection{Technical Literature Review}
There have been many recent works to model the VVLC channel using the radiation pattern of an LED in outdoor scenarios. In \cite{Karbalayghareh:2020}, the vehicle-to-vehicle (V2V) VLC path-loss was modelled with asymmetrical patterns of headlights in various weather conditions using ray tracing simulations. The authors in \cite{Chen:2016} empirically analyzed the temporal variation of V2V VLC channels affected by vehicular movements. A machine learning-based VVLC channel modeling was proposed to improve the path-loss accuracy considering the vehicular mobility and environmental information in \cite{Turan:2021}. In \cite{Eldeeb:2021}, the V2V VLC path-loss model obtained by signals asymmetrically radiated from taillights in ray tracing simulations was studied in comparison with models derived from the measurements and Lambertian radiation pattern. An infrastructure-to-vehicle (I2V) VLC path-loss was represented with geometrical parameters and evaluated through the bit error rate (BER) derived by probability density function of path-loss in \cite{Eldeeb:2022}.

With the predefined radiation pattern of an LED, the transmission area of visible light is limited typically like the Lambertian radiation pattern. The precoding in VLC systems only restricts areas radiating signals by controlling the magnitude of signal from each LED without manually switching direction. A precoding scheme based on the space-time block code with orthogonal circulant matrix transform improved the performance of multiple-input multiple-output (MIMO) orthogonal frequency division multiplexing (OFDM) VLC systems in \cite{Chen:2020}. In \cite{Yosef:2024}, the non-orthogonal multiple access (NOMA) in VVLC systems enhanced the spectral efficiency by using beamforming techniques and the successive interference cancellation (SIC). In \cite{Cespedes:2023}, the blind interference alignment was performed to achieve high data rates and low latency by obtaining linearly independent VVLC channels using a reconfigurable photodiode (PD).

Many previous works have been studied to mitigate the interference by ambient light. In \cite{Memedi:2020}, the sunlight generating direct current (DC) was reduced by low-pass filtering. In \cite{Illanko:2021}, the interference from sunlight could be minimized by optimizing the incident angles of i) transmitted signals and ii) ambient light arrived at the PDs to maximize the signal-to-noise ratio (SNR). The interference and ambient light were spatially filtered by using a liquid crystal panel in \cite{Tebruegge:2019}. The authors in \cite{Kumar:2012} modulated signals with direct sequence spread spectrum (DSSS) and sequence inverse keying (SIK) to minimize the effect of ambient light. A differential optical receiver was developed to minimize the interference from ambient light in \cite{Alam:2016}.

VVLC systems including multiple vehicles have also been investigated to account for practical vehicular environments. In \cite{Yahia:2021}, a receive (Rx) vehicle employed the angular diversity by multiple PDs when receiving signals from headlights of multiple vehicles. In \cite{Plascencia:2023}, the packet delivery ratio (PDR) was evaluated to verify the effect of multi-user interference by multiple taillights based on analytical and experimental approaches. Optical code division multiple access (CDMA) techniques mitigated the multi-user interference to improve PDR performance in \cite{Plascencia2:2023}.

\subsection{Motivation}
While a few works, e.g., \cite{Yosef:2024,Tebruegge_2:2019}, developed precoding methods for VVLC systems, most of prior works on the multi-user VLC precoding design considered indoor scenarios. In \cite{Cho:2019} and \cite{Cho:2020}, the authors i) discussed the simultaneous precoding and jamming to maximize the signal-to-interference-plus-noise ratio (SINR) of a legitimate receiver while suppressing the SINR of an eavesdropper, and ii) proposed the relaxed zero-forcing (ZF) precoding considering both active and passive eavesdroppers, respectively. In \cite{Hong:2013}, the block diagonalization precoding when using PDs with different field of views (FoVs) was analyzed in multi-user MIMO VLC systems. Optimal precoders to maximize the minimum rate and sum rate in multi-user multiple-input single-output (MISO) VLC systems were designed by solving convex optimization problems with the amplitude constraint in \cite{Pham:2017}. In \cite{Pham:2019}, the cell coordination/cooperation strategies and corresponding precoder designs were proposed to mitigate the inter-cell interference in multi-user multi-cell VLC systems.

The precoding methods in multi-user VLC systems studied in these works are based on distinguishable VLC channels thanks to the sufficient spacing between LEDs. VVLC channels, however, are highly correlated as a result of densely positioned LEDs in each headlight. The high correlation forms a nearly identical vector space for VVLC channels in multi-user systems, which degrades the sum rate by causing serious inter-user interference (IUI). A few prior works have studied for adaptive front-lighting system (AFS) to mitigate the IUI by forming the sharp radiation pattern to support multiple vehicles simultaneously \cite{Memedi:2020,Tebruegge_2:2019}. This approach, however, does not fully use available resources and may not work for Rx vehicles that are far from a transmit (Tx) vehicle but located close to each other due to the radiation range of each LED chip itself. Furthermore, the IUI mitigation using AFS can be performed only when headlights have a large number of LEDs, which may cause higher computational complexity to choose LEDs to suppress the IUI. To enhance the practicality of IUI mitigation, we need a novel method to decorrelate the LEDs so that the VVLC channels span distinct vector spaces.

\subsection{Our Contributions}
The precoder design in multi-user VVLC systems equipped with gold nanoparticles (GNPs) can be a solution to achieve high data rates by reducing the correlation between the LEDs. GNPs, recently synthesized metamaterials, have chiroptical properties exhibiting significantly different absorption and phase delay for left/right circularly polarized light \cite{Lee:2018}. These properties vary depending on the wavelength and azimuth angle of incident light. The transmittance obtained from i) the differential absorption effect according to the azimuth angle of incident light and ii) different cell arrangement in a GNP plate can decorrelate LEDs by making the row spaces of the VVLC channels be independent, which enables a Tx vehicle to send independent data to multiple Rx vehicles. After decorrelating LEDs with GNPs, we adopt a signal-to-leakage-plus-noise ratio (SLNR)-based precoder to suppress IUI between multiple Rx vehicles in dynamic vehicular environments.

Our contributions in this paper are summarized as follows.
\begin{itemize}
\item We first design a multi-user MIMO VVLC system using the differential absorption effect depending on the azimuth angle of incident light and the elaborate cell arrangement in a GNP plate to reduce the correlation between LEDs.
\item The SLNR-based precoders are obtained using the generalized Rayleigh quotient based on the approximation of shot noise that is proportional to the received signal power.
\item We generalize the color mapping method on the CIE 1931 $xy$ coordinates of IEEE standard 802.15.7 to derive a white color constraint when using multiple LEDs having RGB light sources.
\item The ratios of RGB light sources are optimized by maximizing the sum SLNR with the successive convex approximation (SCA) method under a relaxed white light constraint.
\item The simulation results validate the high correlation of VVLC channels and justify the half power angle setting of LED. Moreover, the proposed SLNR-based precoder design considerably enhances the sum rate in a multiple access scenario and the secrecy rate in a wiretapping scenario.
\end{itemize}

There have been several recent works to verify the feasibility of VVLC systems by experiments \cite{Tettey:2023,Yang:2023,Aly:2023,Al:2025}. We, however, verify the effectiveness of proposed SLNR-based precoding method in multi-user VVLC systems by simulations. The dynamic color modulation using GNPs was tested by real experiments in indoor scenarios in \cite{Lee:2018}, and optical measurements on GNPs were carried out under various configurations \cite{Kim:2024,Kim:2023capacitive,Ohnoutek:2020,Skvortsova:2024}, which support the robustness of GNPs across different optical platforms. In addition, GNPs exhibit remarkable resistance to outdoor environments such as dust, aging, and both mechanical and thermal stress. In particular, gold's the highest standard reduction potential leads to exceptional resistance to oxidation and degradation caused by environmental exposure, and embedding GNPs in a polymer template further enhances their environmental stability \cite{Li:2023,Liu:2009,Kim:2011}. Based on these findings, we expect that GNP-empowered VLC systems are possible for outdoor applications, e.g., vehicular environments.

In addition, we believe our proposed GNP-empowered VVLC system will become highly practical in the near future due to several reasons.

\textit{Reason 1:} As verified in \cite{Han:2024}, the fabrication cost of GNPs is very low, making their mass production possible. Also, a GNP plate can be produced at a low cost, further facilitating the mass production \cite{Kim:2022}.

\textit{Reason 2:} Even though legacy LEDs such as reflector and projector LEDs are typically used for vehicular headlights to illuminate surroundings, the RGB LEDs, which can dynamically adapt to proper color temperature and brightness for driving environments, are also already commercialized \cite{RGB_LED_1,RGB_LED_2}. The RGB LED is very useful to improve VVLC performance since it can provide additional degree of freedom (DoF), e.g., wavelength of light. Thus, future vehicles communicating with each other using visible spectrum would widely adopt the RGB LEDs to achieve high performance in communications while ensuring stable illumination.

\textit{Reason 3:} The commercial LEDs today are typically connected by a shared driver since they do not need to modulate any signals to perform only a lighting purpose, which makes the independent modulation difficult. The AFS with the matrix LED headlight, however, recently demonstrated the feasibility of independent LED control to prevent the glare for oncoming vehicles \cite{Kim:2023,Cerri:2019}. In \cite{Bai:2024}, the author proposed the LED matrix managers to develop the intelligent and safe front light system by dimming LEDs individually. Also, VLC systems using LEDs, like antennas in RF systems, should modulate signals independently to effectively transmit data while simultaneously illuminating surroundings. In \cite{Yosef:2024,Tebruegge_2:2019,Schettler:2019}, the AFS has been adopted in VVLC systems to enable the spatial multiplexing while reducing the IUI. Even though widely used headlights today may not fully support per-LED modulation, the proposed SLNR-based precoding in this paper aligns with emerging trends using smart headlight designs, where LEDs can be individually controlled for adaptive lighting and communications.

The remainder of paper is organized as follows. In Section \ref{sec_system}, we first present the chiroptical properties of GNPs, derive the effective VVLC channel integrating the effect of geometric path-loss, GNPs, and RGB light sources, and then represent the received signal model. We optimize the SLNR-based precoder by approximating the shot noise with arbitrary RGB ratios in Section \ref{sec_pre}. In Section \ref{sec_RGB_ratio}, we obtain the white light constraint for headlights having multiple LEDs by generalizing the color mapping method of IEEE standard 802.15.7, relax it to ensure a larger feasible set, and finally obtain the optimal RGB ratios less absorbed by GNPs employing the SCA method to maximize the sum SLNR. In Section \ref{sec_AO}, we provide an algorithm for updating the SLNR-based precoders and RGB ratios alternatingly to maximize the sum SLNR and analyze the complexity of proposed algorithm. In Section \ref{sec_simul}, we discuss the performance improvement by the proposed precoder design with the chromaticity values of optimized RGB ratios via simulation results. Finally, we conclude this paper in Section \ref{sec_conc}.

\begin{figure}
	\centering
	\includegraphics[width=1.0\columnwidth]{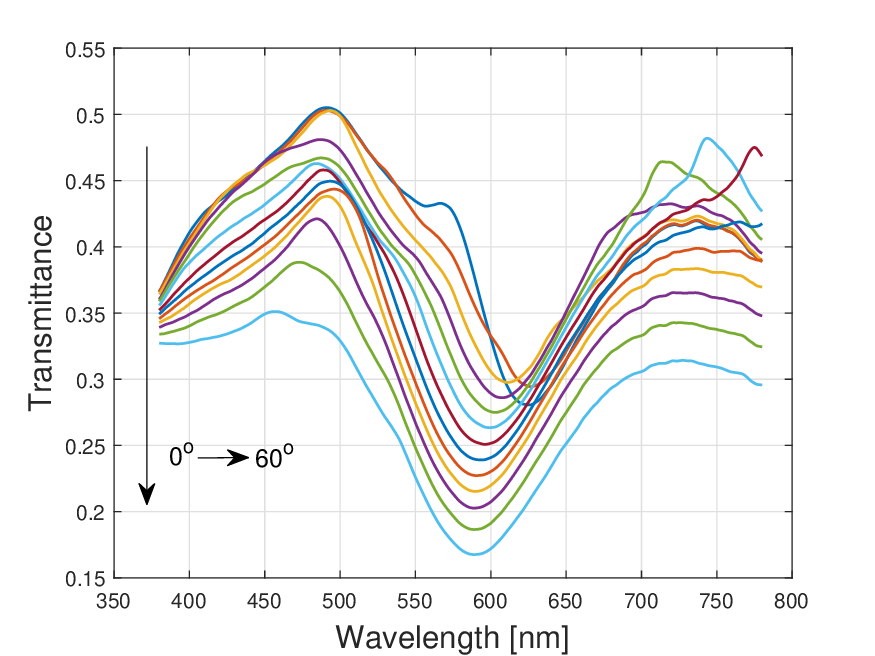}
	\caption{Transmittance by one cell of a GNP plate depending on the azimuth angle of incident light with the step size of $5^{\circ}$.}\label{Trans_GNP}
\end{figure}

\textit{Notation:} $a$, $\ba$, and $\bA$ denote a scalar, vector, and matrix. $|a|$ denotes the magnitude of $a$. $\|\ba\|$ represents the Euclidean norm of $\ba$. $(\cdot)^{\mathrm{T}}$, $(\cdot)^*$, and $(\cdot)^{-1}$ are the transpose, complex conjugate, and inverse. A normal distribution with mean $\mu$ and variance $\sigma^2$ is denoted as $\mathcal{N}(\mu,\sigma^2)$. $\mathbb{R}_{+}$ represents the set of positive real numbers. $\mathcal{\mathbf{1}}_N$, $\mathbf{0}_{N}$, $\mathcal{\mathbf{1}}_{M\times N}$, $\mathcal{\mathbf{0}}_{M\times N}$, and $\bI_N$ denote the $N\times 1$ all-ones and all-zeros vectors, all-ones and all-zeros matrices with the size $M\times N$, and identity matrix with the size $N\times N$. The Kronecker product of $\ba$ and $\bb$ is represented as $\ba \otimes \bb$. $\mathrm{max}(a,b)$ is the max function that returns the largest value between $a$ and $b$. $\mathrm{diag}(\cdot)$ denotes the diagonal matrix. $\mathbb{E}\left\{\cdot\right\}$ is the expectation operator.

\section{System Model}\label{sec_system}
\subsection{Chiroptical Properties of GNPs}
The chiroptical properties of GNPs exhibit circular dichroism (CD) and optical rotatory dispersion (ORD) for incident light, which manifests differential absorption and refraction of left/right circularly polarized light \cite{Lee:2018}. The left/right circularly polarized light experiences different amounts of phase delay by the differential refraction. These chiroptical properties themselves can be varied by the type, pattern, and size of GNPs \cite{Han:2024,Kim:2021}. The large size of GNPs makes the chiroptical properties such as CD and ORD to be red-shifted due to the variation of wavelength that leads to the localized surface plasmon resonance (LSPR) depending on the diffraction order \cite{Kim:2021,Kim:2022}, which is similar to the different type of GNPs \cite{Lee:2018,Cho:2022,Kim:2020_gamma,Im:2023}. The hexagonal pattern also has the red-shifted chiroptical properties compared to the square pattern because the wavelength that induces the lattice resonance changes \cite{Kim:2024}. These various configurations of GNPs can improve the sum SLNR by better separating the row spaces of VVLC channels for multiple Rx vehicles, which eventually enhances the communication performances. We can exploit the chiroptical properties depending on the various configurations by controlling polarization, wavelength, and azimuth angle of incident light for designing the VVLC system with GNPs.

In this paper, however, the polarization is not employed since it reduces the Tx power from LEDs, which may reduce communication distance. As neither the transmitter nor the receiver is equipped with a polarizer, the proposed framework does not exploit the phase delay effect of GNPs, which is only observable from variations in the polarization state and otherwise averaged out as described in ``Supplementary Information'' of \cite{Lv:2024}. Instead, we employ the differential transmittance depending on the wavelength and azimuth angle of incident light as depicted in Fig. \ref{Trans_GNP}. A GNP plate is fabricated by stacking GNPs with different types, patterns, and sizes \cite{Lee:2018}. By the elaborated arrangement of GNPs, a GNP plate can consist of several cells with different chiroptical properties assuming all areas in a cell have the same properties. We assume the signals radiated from an LED pass through only one cell among $N$ cells of a GNP plate.

\textit{Remark:} Note that we exploit the GNP plate to distinguish the signals for multiple Rx vehicles instead of relying on wavelength division multiplexing (WDM). WDM requires many optical filters to separate signals carried by different wavelengths and may suffer from crosstalk by the wide spectral power distribution (SPD) of light source. On the contrary, the GNP plate can separate the signals by the chiroptical properties depending on the azimuth angle of incident light without using any optical filters. This would significantly lower the burden on the receivers that typically perform huge amounts of signal processing for WDM. Especially, in our proposed GNP-empowered VVLC system, the Rx vehicles do not suffer from any crosstalk.

\begin{figure}
	\centering
	\includegraphics[width=1.0\columnwidth]{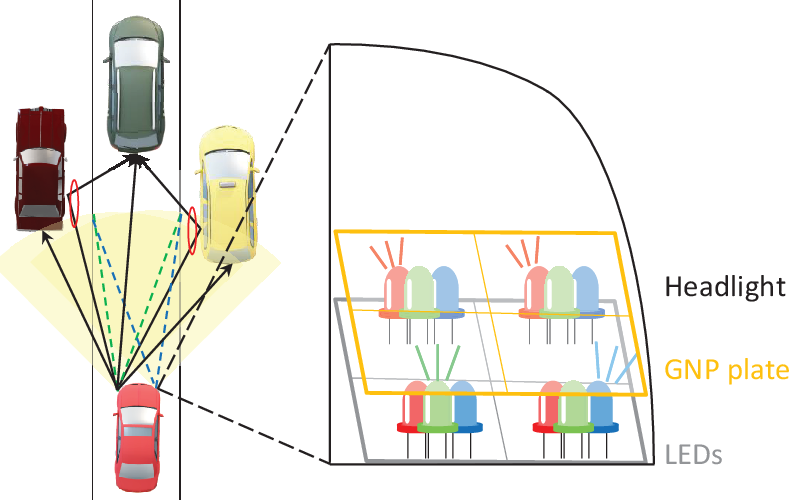}
	\caption{A multi-user MIMO VVLC system considered in this paper.}\label{MA_scenario}
\end{figure}

\subsection{Effective Channel Model}
We consider a multi-user MIMO VVLC system with $N_{\mathrm{t}}$ LEDs and a GNP plate in each headlight of Tx vehicle and $N_{\mathrm{r}}$ PDs at the center of each Rx vehicle's rear-end, as shown in Fig. \ref{MA_scenario}. Each cell in a GNP plate has different transmittance, and to design independent precoders for $U$ Rx vehicles, we assume $N\ge U$. The transmitted signals from $N_{\mathrm{t}}$ LEDs of the $i$-th headlight to support $U$ Rx vehicles are given as
\setcounter{equation}{0}
\begin{equation}
\bx_i=P_{\mathrm{Tx}}\mathbf{1}_{N_{\mathrm{t}}}+\alpha P_{\mathrm{Tx}}\sum_{u=1}^{U}\mathbf{f}_{u}^i s_u^i,
\end{equation}
where $P_{\mathrm{Tx}}=\zeta I_{\mathrm{DC}}$ is the Tx power with the current-to-light conversion efficiency $\zeta$ and the DC bias for illumination $I_{\mathrm{DC}}$, $\alpha\in [0,1]$ is the modulation index, $\mathbf{f}_u^i$ and $s_u^i$ are the precoder and pulse amplitude modulation (PAM) constellation symbol with $\mathbb{E}\left\{ s_u^i \right\} = 0$ and $\mathbb{E}\left\{ | s_u^i |^2 \right\} = 1$ from the $i$-th headlight to the $u$-th Rx vehicle, and $i\in\left\{ 1,2 \right\}$.

\setcounter{equation}{9}
\begin{figure*}[t]
	\centering
	\vspace{-\baselineskip}
	\vspace{-\baselineskip}
	\begin{align}\label{eq10}
	&\bT( \bN(\lambda,\phi) \otimes \bN(\lambda,\phi)^* )\bT^{-1} = \frac{1}{2}\begin{bmatrix}
        \bN_1(\lambda,\phi) & \b0_{2\times 2} \\
        \b0_{2\times 2} & \bN_2(\lambda,\phi)
	\end{bmatrix},\notag\\
        & \bN_1(\lambda,\phi) = \begin{bmatrix}
        \bar{a}_{\mathrm{L}}(\lambda,\phi) + \bar{a}_{\mathrm{R}}(\lambda,\phi) & \bar{a}_{\mathrm{L}}(\lambda,\phi) - \bar{a}_{\mathrm{R}}(\lambda,\phi) \\
        \bar{a}_{\mathrm{L}}(\lambda,\phi) - \bar{a}_{\mathrm{R}}(\lambda,\phi) & \bar{a}_{\mathrm{L}}(\lambda,\phi) + \bar{a}_{\mathrm{R}}(\lambda,\phi), \\
        \end{bmatrix},\notag\\
        & \bN_2(\lambda,\phi) = 2\sqrt{ \bar{a}_{\mathrm{L}}(\lambda,\phi) \bar{a}_{\mathrm{R}}(\lambda,\phi) } \begin{bmatrix}
        \cos(\Delta\varphi(\lambda,\phi)) & -\sin(\Delta\varphi(\lambda,\phi)) \\
        \sin(\Delta\varphi(\lambda,\phi)) & \cos(\Delta\varphi(\lambda,\phi))
        \end{bmatrix}.
        \end{align}
	\hrule
\end{figure*}

The effective VVLC channel consists of geometric path-loss and the effect of GNPs and RGB light sources as
\setcounter{equation}{1}
\begin{equation}\label{eq2}
\tilde{\bH}_u^i = \bH_u^i \diag( \bG_u^i \bp^i )
\end{equation}
with
\begin{align}\label{eq3}
\bG_u^i &= \begin{bmatrix}
(\bg_{u,1}^i)^{\mathrm{T}} & & \\
& \ddots & \\
& &  (\bg_{u,N_{\mathrm{t}}}^i)^{\mathrm{T}}
\end{bmatrix}\in \mathbb{R}_{+}^{ N_{\mathrm{t}} \times 3N_{\mathrm{t}} },\notag\\
\bp^i &= \begin{bmatrix}
( \bp_1^i )^{\mathrm{T}} & \cdots & ( \bp_{N_{\mathrm{t}}}^i )^{\mathrm{T}}
\end{bmatrix}^{\mathrm{T}}\in\mathbb{R}_{+}^{3N_{\mathrm{t}}\times 1},
\end{align}
where $\bH_u^i\in \mathbb{R}_{+}^{N_{\mathrm{r}}\times N_{\mathrm{t}}}$ is the geometric path-loss according to the Lambertian radiation pattern\footnote{While we do not consider different weather conditions in this paper, it would be interesting to investigate the impact of weather for our proposed technique as in \cite{Karbalayghareh:2020}.}, and $\bg_{u,m}^i\in \mathbb{R}_{+}^{3\times 1}$ and $\bp_m^i\in\mathbb{R}_{+}^{3\times 1}$ denote the transmittance by GNPs and the ratio for RGB light sources in each LED, respectively.

\subsubsection{Geometric path-loss}
The $(n,m)$-th component of $\bH_u^i$, $h_{u,nm}^i$, is the geometric path-loss from the $m$-th LED at the $i$-th headlight of Tx vehicle to the $n$-th PD at the $u$-th Rx vehicle. There are line-of-sight (LOS) and non-LOS paths in $\bH_u^i$ as \cite{Lee:2011}
\setcounter{equation}{3}
\begin{equation}
h_{u,nm}^i = ( h_{u,nm}^i )_{\mathrm{L}} + ( h_{u,nm}^i )_{\mathrm{NL}}
\end{equation}
with
\begin{align}\label{eq5}
(h_{u,nm}^i)_{\mathrm{L}} &= \frac{ ( n_{\mathrm{LED}} + 1 )A_{\mathrm{PD}} }{ 2\pi R_{\mathrm{L},nm}^2 }\cos^{n_{\mathrm{LED}}}\phi_{\mathrm{L},nm}\cos\theta_{\mathrm{L},nm},\notag\\
(h_{u,nm}^i)_{\mathrm{NL}} &= \left( \frac{ ( n_{\mathrm{LED}} + 1 )A_{\mathrm{ref}}\cos^{n_{\mathrm{LED}}}\phi_{\mathrm{NL},m}\cos\theta_{\mathrm{NL},m} }{ 2\pi R_{\mathrm{NL},m}^2 } \right)\notag\\
&\qquad\qquad\qquad \times\left( \frac{ A_{\mathrm{PD}} \cos \phi_{\mathrm{NL},n}\cos\theta_{\mathrm{NL},n} }{ \pi R_{\mathrm{NL},n}^2 } \bar{\rho} \right).
\end{align}
In (\ref{eq5}), 
\begin{equation}\label{eq6}
\bar{\rho} = \frac{ \int_{\lambda} S(\lambda)\rho(\lambda) d\lambda }{ \int_{\lambda} S(\lambda) d\lambda }
\end{equation}
is the average reflectance with the spectral reflectance of Rx vehicle $\rho(\lambda)$ and the SPD of LED $S(\lambda)=\sum_{c=1}^{3} S_c(\lambda)$ composed of the sum of SPDs for the RGB light sources $S_c(\lambda)$, $n_{\mathrm{LED}} = \frac{ -\ln 2 }{ \ln \cos ( \Phi_{1/2} ) }$ is the Lambertian emission order with the half power angle of LED $\Phi_{1/2}$, $A_{\mathrm{PD}}$ and $A_{\mathrm{ref}}$ denote the areas of PD and reflector, $R_{\mathrm{L},nm}$ and $(\phi_{\mathrm{L},nm},\theta_{\mathrm{L},nm})$ are the distance and the radiation and incidence angle pair of the LOS path, $(R_{\mathrm{NL},m},R_{\mathrm{NL},n})$ and $\{ (\phi_{\mathrm{NL},m},\theta_{\mathrm{NL},m}), (\phi_{\mathrm{NL},n},\theta_{\mathrm{NL},n})\}$ are the distances and the radiation and incidence angle pairs of the non-LOS paths from the $m$-th LED to the reflector and from the reflector to the $n$-th PD, respectively. The indices of RGB light sources for $c=1,2,3$ represent red, green, and blue, respectively.

\subsubsection{Effect of GNPs and RGB light sources}
The chiroptical properties of GNPs can be represented as the Jones matrix in the circular polarization domain as
\setcounter{equation}{6}
\begin{align}\label{eq7}
\bN(\lambda,\phi) = \begin{bmatrix}
\sqrt{\bar{a}_{\mathrm{L}}(\lambda,\phi)} & 0 \\
0 & \sqrt{\bar{a}_{\mathrm{R}}(\lambda,\phi)} e^{j\Delta\varphi(\lambda,\phi)}
\end{bmatrix},
\end{align}
where $(\bar{a}_{\mathrm{L}}(\lambda,\phi),\bar{a}_{\mathrm{R}}(\lambda,\phi))$ and $\Delta\varphi(\lambda,\phi)$ are the transmittance pair and differential phase delay for left/right circularly polarized light with the wavelength $\lambda$ and azimuth angle $\phi$. To represent the transmittance by GNPs for unpolarized light, we transform the Jones matrix in (\ref{eq7}) into the M\"uller matrix as
\begin{align}
\bM = \bT ( \bJ\otimes \bJ^* ) \bT^{-1}
\end{align}
with
\begin{align}
\bT = \begin{bmatrix}
1 & 0 & 0 & 1 \\
1 & 0 & 0 & -1 \\
0 & 1 & 1 & 0 \\
0 & j & -j & 0
\end{bmatrix},
\end{align}
where $\bJ$ is an arbitrary Jones matrix. The M\"uller matrix of GNPs can be denoted as (\ref{eq10}) at the top of this page. The Stokes vector describes all the polarization state as
\setcounter{equation}{10}
\begin{align}
\begin{bmatrix}
S_1 \\ S_2 \\ S_3 \\ S_4
\end{bmatrix}=
\begin{bmatrix}
I \\
Ip\cos(2\chi_1)\cos(2\chi_2) \\
Ip\sin(2\chi_1)\cos(2\chi_2) \\
Ip\sin(2\chi_2)
\end{bmatrix},
\end{align}
where $I$ is the intensity of light, $p$ is the degree of polarization, and $(\chi_1,\chi_2)$ are the ORD and CD. Finally, the transmittance by GNPs for unpolarized light can be obtained from the first component of Stokes vector after passing through GNPs as $\frac{ \bar{a}_{\mathrm{L}}(\lambda,\phi) + \bar{a}_{\mathrm{R}}(\lambda,\phi) }{2}$.

Note that, without GNPs, the transmitted signals experience nearly identical channels with just different path-loss. The high correlation between LEDs, and eventually the high correlation of channels for multiple Rx vehicles, can be lowered by the different transmittance thanks to i) the differential absorption effect of GNPs and ii) cell arrangement with different types, patterns, and sizes of GNPs in a GNP plate.

The $c$-th component of $\bg_{u,m}^i$ can be represented as
\setcounter{equation}{11}
\begin{align}
g_{u,m,c}^i = \int_{\lambda} R_{\mathrm{PD}}(\lambda) S_{c}(\lambda) \bar{a}_{u,m}^i(\lambda) d\lambda,
\end{align}
for $c\in\left\{ 1,2,3 \right\}$, where $R_{\mathrm{PD}}(\lambda)$ is the responsivity of PD, and $\bar{a}_{u,m}^i(\lambda) = \frac{\bar{a}_{\mathrm{L},u,m}^i(\lambda) + \bar{a}_{\mathrm{R},u,m}^i(\lambda)}{2}$ is the transmittance by GNPs with the transmittance for left/right circularly polarized light $(\bar{a}_{\mathrm{L},u,m}^i(\lambda),\bar{a}_{\mathrm{R},u,m}^i(\lambda))$. We use the index of Rx vehicle $u$ instead of azimuth angle of incident light in the transmittance by GNPs.

\subsection{Received Signal Model}
The received signal model at the $u$-th Rx vehicle after eliminating the DC bias is given as
\begin{equation}
\by_u = \alpha P_{\mathrm{Tx}}\sum_{i=1}^{2}\left( \tilde{\bH}_u^i \mathbf{f}_u^i s_u^i + \sum_{v\ne u} \tilde{\bH}_u^i \mathbf{f}_v^i s_v^i \right) + \bn_u,
\end{equation}
where $\bn_u \sim \mathcal{N}(0,\sigma_u^2)$ is the noise at the $u$-th Rx vehicle with the variance $\sigma_u^2 = \sigma_{\mathrm{th}}^2 + \sum_{i=1}^{2} \sigma_{\mathrm{sh},u,i}^2$ with the thermal noise $\sigma_{\mathrm{th}}^2$ and the shot noise at the $u$-th Rx vehicle by the $i$-th headlight $\sigma_{\mathrm{sh},u,i}^2$. The shot noise becomes dominant compared to the thermal noise for high SNR regions since it is proportion to the received power as
\begin{equation}
\sigma_{\mathrm{sh},u,i}^2 = \gamma P_{\mathrm{Tx}} \mathbb{E} \left\{ \begin{Vmatrix}
\tilde{\bH}_u^{i} ( \mathbf{1}_{N_{\mathrm{t}}} + \alpha \sum_{v=1}^{U} \mathbf{f}_v^{i} s_v^{i} ) \end{Vmatrix} \right\},
\end{equation}
where $\gamma = 2qB$ with the electronic charge $q$ and the system bandwidth $B$ \cite{Sindhubala:2017,Liu:2020sh}. Thus, the shot noise needs to be considered for designing precoders in VVLC systems primarily driving on the high power range.

\section{SLNR-Based Precoder Design}\label{sec_pre}
We first explain the proposed SLNR-based precoder assuming the RGB ratio $\bp^i$, which determines the effective channel $\tilde{\bH}_u^i$ in (\ref{eq2}), is fixed. In Section \ref{sec_RGB_ratio}, we elaborate how to design $\bp^i$ to maximize the sum SLNR. We further assume that the Tx vehicle only knows the LOS path for the precoder design and the RGB ratio optimization since the Tx vehicle may not be able to estimate the non-LOS paths using sensors equipped on it in practice. In Section \ref{sec_simul}, we numerically verify the effect of non-LOS paths on sum rates and secrecy rates.

\setcounter{equation}{19}
\begin{figure*}[t]
	\centering
	\vspace{-\baselineskip}
	\vspace{-\baselineskip}
	\begin{align}\label{eq20}
	x_{u}^i&=\frac{ \sum_{m=1}^{N_{\mathrm{t}}}\sum_{c=1}^3 P_{m,c}^i X_{u,m,c}^i }{ \sum_{m=1}^{N_{\mathrm{t}}}\sum_{c=1}^3 P_{m,c}^i \left( X_{u,m,c}^i + Y_{u,m,c}^i + Z_{u,m,c}^i \right) } = \frac{ \begin{bmatrix} ( \bt_{X,u,1}^i )^{\mathrm{T}} & \cdots & ( \bt_{X,u,N_{\mathrm{t}}}^i )^{\mathrm{T}} \end{bmatrix} \bp^i }{ \begin{bmatrix} ( \bt_{T,u,1}^i )^{\mathrm{T}} & \cdots & ( \bt_{T,u,N_{\mathrm{t}}}^i )^{\mathrm{T}} \end{bmatrix} \bp^i } = \frac{ ( \bt_{X,u}^i )^{\mathrm{T}} \bp^i }{ ( \bt_{T,u}^i )^{\mathrm{T}} \bp^i }, \notag\\
	y_{u}^i&=\frac{ \sum_{m=1}^{N_{\mathrm{t}}}\sum_{c=1}^3 P_{m,c}^i Y_{u,m,c}^i }{ \sum_{m=1}^{N_{\mathrm{t}}}\sum_{c=1}^3 P_{m,c}^i \left( X_{u,m,c}^i + Y_{u,m,c}^i + Z_{u,m,c}^i \right) } = \frac{ \begin{bmatrix} ( \bt_{Y,u,1}^i )^{\mathrm{T}} & \cdots & ( \bt_{Y,u,N_{\mathrm{t}}}^i )^{\mathrm{T}} \end{bmatrix} \bp^i }{ \begin{bmatrix} ( \bt_{T,u,1}^i )^{\mathrm{T}} & \cdots & ( \bt_{T,u,N_{\mathrm{t}}}^i )^{\mathrm{T}} \end{bmatrix} \bp^i } = \frac{ ( \bt_{Y,u}^i )^{\mathrm{T}} \bp^i }{ ( \bt_{T,u}^i )^{\mathrm{T}} \bp^i },\notag\\
    \bt_{X,u,m}^i &= [ X_{u,m,1}^i, \cdots, X_{u,m,3}^i ]^{\mathrm{T}}, \quad
    \bt_{Y,u,m}^i = [ Y_{u,m,1}^i, \cdots, Y_{u,m,3}^i ]^{\mathrm{T}},\notag\\
    \bt_{T,u,m}^i &= [ X_{u,m,1}^i+Y_{u,m,1}^i+Z_{u,m,1}^i, \cdots, X_{u,m,3}^i+Y_{u,m,3}^i+Z_{u,m,3}^i ]^{\mathrm{T}}.
	\end{align}
	\hrule
\end{figure*}

In typical multi-user MIMO systems, the SLNR maximization can be addressed using a function of generalized Rayleigh quotient as in \cite{Sadek:2007}. In VLC systems, the SLNR maximization is not the form of generalized Rayleigh quotient due to the shot noise, whose upper bound having a closed form is derived by the Jensen's inequality as
\setcounter{equation}{14}
\begin{align}\label{eq15}
\sigma_{\mathrm{sh},u,i}^2 &= \gamma P_{\mathrm{Tx}} \mathbb{E} \left\{ \begin{Vmatrix}
\tilde{\bH}_u^{i} ( \mathbf{1}_{N_{\mathrm{t}}} + \alpha \sum_{v=1}^{U} \mathbf{f}_v^{i} s_v^{i} ) \end{Vmatrix} \right\}\notag\\
&\le \gamma P_{\mathrm{Tx}} \sqrt{ \| \tilde{\bH}_u^{i}\mathbf{1}_{N_{\mathrm{t}}} \|^2 + \alpha^2 \sum_{v=1}^{U} \| \tilde{\bH}_u^{i} \mathbf{f}_v^{i} \|^2 }\notag\\
&= \tilde{\sigma}_{\mathrm{sh},u,i}^2,
\end{align}
by using the fact $\mathbb{E}\left\{ | s_v^i |^2 \right\} = 1$. Even though $\tilde{\sigma}_{\mathrm{sh},u,i}^2$ is a function of precoders that are the variables of following optimization problem, we aim to make $\tilde{\sigma}_{\mathrm{sh},u,i}^2$ to be a constant. This ensures that the SLNR becomes the form of generalized Rayleigh quotient, which is favorable since the generalized Rayleigh quotient maximization yields a straightforward solution. Thanks to high correlation by the very small gap between PDs in each Rx vehicle, the rank of $\tilde{\bH}_u^{i}$ becomes nearly one even though the GNPs decorrelate LEDs in each headlight. The product of effective VVLC channels can be approximated as
\begin{equation}\label{eq16}
(\tilde{\bH}_u^{i})^{\mathrm{T}}\tilde{\bH}_u^{i}\approx \tilde{\lambda}_u^{i}\tilde{\bq}_u^{i}(\tilde{\bq}_u^{i})^{\mathrm{T}},
\end{equation}
with the maximum eigenvalue $\tilde{\lambda}_u^{i}$ and the eigenvector $\tilde{\bq}_u^i$ corresponding to $\tilde{\lambda}_u^{i}$. In Section \ref{sec_simul}, the simulation result for the condition number of $\tilde{\bH}_u^{i}$ depending on the distance verifies that the approximation in (\ref{eq16}) is reasonable. To remove the impact of precoders, $\tilde{\sigma}_{\mathrm{sh},u,i}^2$ can be further bounded as
\begin{align}
\tilde{\sigma}_{\mathrm{sh},u,i}^2 &\stackrel{(a)}{\approx} \gamma P_{\mathrm{Tx}} \sqrt{ \tilde{\lambda}_u^{i} } \cdot \sqrt{ ( \mathbf{1}_{N_{\mathrm{t}}}^{\mathrm{T}} \tilde{\bq}_u^{i} )^2 + \alpha^2 \sum_{v=1}^U \left( ( \mathbf{f}_v^{i} )^{\mathrm{T}} \tilde{\bq}_u^{i}  \right)^2 } \notag\\
&\stackrel{(b)}{\le} \gamma P_{\mathrm{Tx}} \sqrt{ \tilde{\lambda}_u^{i} } \left( \begin{vmatrix} \mathbf{1}_{N_{\mathrm{t}}}^{\mathrm{T}}\tilde{\bq}_u^{i}
\end{vmatrix} + \alpha \sum_{v=1}^U \begin{vmatrix}
( \mathbf{f}_v^{i} )^{\mathrm{T}}\tilde{\bq}_u^{i} 
\end{vmatrix} \right) \notag\\
&\stackrel{(c)}{\le} \gamma P_{\mathrm{Tx}} \sqrt{ \tilde{\lambda}_u^{i} } \left( \| \mathbf{1}_{N_{\mathrm{t}}} \| + \alpha \sum_{v=1}^U \| \mathbf{f}_v^{i} \| \right) \notag\\
&= \gamma P_{\mathrm{Tx}} ( \sqrt{N_{\mathrm{t}}} + \alpha U ) \sqrt{ \tilde{\lambda}_u^{i} }\notag\\
&= \check{\sigma}_{\mathrm{sh},u,i}^2,
\end{align}
where $(a)$ comes from the approximation in (\ref{eq16}), $(b)$ is from the triangle inequality, and $(c)$ is due to the Cauchy-Schwarz inequality. Finally, the SLNR maximization problem can be represented as
\begin{align}
&\mathrm{(P1):}\,\, \underset{\mathbf{f}_u^i}{\max}\,\,\,\frac{ \alpha^2 P_{\mathrm{Tx}}^2(\mathbf{f}_u^i)^{\mathrm{T}} (\tilde{\bH}_u^i)^{\mathrm{T}}\tilde{\bH}_u^i \mathbf{f}_u^i }{ (\mathbf{f}_u^i)^{\mathrm{T}} \left( \bL_u^i + ( \check{\sigma}_{\mathrm{sh},u,i}^2 + \sigma_{\mathrm{th}}^2 ) \bI_{N_{\mathrm{t}}} \right) \mathbf{f}_u^i } \notag\\
&\qquad\quad\,\,\, \mathrm{s.t.}\quad \|\mathbf{f}_u^i \| = 1,\tag{1-a}
\end{align}
for $u\in \{ 1,\cdots,U \}$, $i\in\{ 1, 2 \}$, where $\bL_u^i = \alpha^2 P_{\mathrm{Tx}}^2 \left( \sum_{v\ne u} (\tilde{\bH}_v^i)^{\mathrm{T}} \tilde{\bH}_v^i \right)$. The optimized precoder $\hat{\mathbf{f}}_u^i$ is the eigenvector corresponding to the maximum eigenvalue of $\left( \bL_u^i + ( \check{\sigma}_{\mathrm{sh},u,i}^2 + \sigma_{\mathrm{th}}^2 ) \bI_{N_{\mathrm{t}}} \right)^{-1}(\tilde{\bH}_u^i)^{\mathrm{T}}\tilde{\bH}_u^i$.

\section{RGB Ratio Optimization}\label{sec_RGB_ratio}
In this section, we first obtain a white light constraint by generalizing the color mapping method on the CIE 1931 $xy$ coordinates when using multiple LEDs with the RGB light sources, and then relax it to have a larger feasible set to search for a better objective value. Finally, we solve the nonconvex optimization problem for the RGB ratio under the relaxed white light constraint by using the quadratic transform and SCA method.

\subsection{White light constraint}\label{white_const}
The color of radiated light passing through GNPs is varied due to the differential absorption depending on the wavelength. Thus, the transmitted signals from each headlight need to be elaborately adjusted to satisfy the white light constraint after passing through the GNP plate. The white light constraint is important for VVLC systems since the white light has broader spectra than the light with a specific wavelength, making it possible to achieve the stable illumination by providing sufficient reflectance across a variety of surface characteristic. Also, the human eye is particularly sensitive to red, green, and blue light. Since the white light stimulates all three types of photoreceptors simultaneously, it makes the illumination appear brighter to drivers. At the same time, the sum of RGB ratios should be one to satisfy the constant unit power from an LED. In \cite{802.15.7}, the IEEE standard 802.15.7 considers just a set of RGB light sources when mapping a color on the CIE 1931 $xy$ coordinates as
\setcounter{equation}{17}
\begin{align}
x = \sum_{c=1}^3 P_c x_c,\quad
y = \sum_{c=1}^3 P_c y_c,
\end{align}
with the sum RGB ratio constraint, i.e., $\sum_{c=1}^{3} P_c = 1$, where $P_c$ is the ratio of $c$-th light source, and the chromaticity values $(x_c,y_c)$ of $c$-th light source are given as
\begin{align}\label{eq19}
x_c = \frac{ X_c }{ X_c + Y_c + Z_c }, \quad y_c = \frac{ Y_c }{ X_c + Y_c + Z_c },
\end{align}
by normalizing the tri-stimulus values $(X_c,Y_c,Z_c)$.

In each headlight of vehicles, however, there exist multiple LEDs to brighten surroundings, and the transmitted light from multiple LEDs makes one color altogether. We generalize the color mapping method to use for VLC systems with multiple LEDs. The CIE 1931 $xy$ coordinates of an arbitrary target color with chromaticity values $(x_u^i,y_u^i)$ by the generalized color mapping can be represented as (\ref{eq20}) at the top of this page, where $P_{m,c}^i$ is the $c$-th component of $\bp_m^i$. In (\ref{eq20}), $(X_{u,m,c}^i$, $Y_{u,m,c}^i$, $Z_{u,m,c}^i)$ are the tri-stimulus values considering the effect of GNPs, which are given as
\setcounter{equation}{20}
\begin{align}
X_{u,m,c}^i&= \int_{\lambda} S_{c}(\lambda) \bar{a}_{u,m}^i(\lambda)\bar{x}(\lambda) d\lambda, \notag\\
Y_{u,m,c}^i&= \int_{\lambda} S_{c}(\lambda)\bar{a}_{u,m}^i(\lambda)\bar{y}(\lambda) d\lambda, \notag\\
Z_{u,m,c}^i&= \int_{\lambda} S_{c}(\lambda)\bar{a}_{u,m}^i(\lambda)\bar{z}(\lambda) d\lambda,
\end{align}
where $\left\{ \bar{x}(\lambda),\bar{y}(\lambda),\bar{z}(\lambda) \right\}$ are the color matching functions denoting the chromatic response of the CIE standard observer. We can analyze (\ref{eq20}) as a method to determine chromaticity values of the radiated light by comprehensively considering the ratios of $3N_{\mathrm{t}}$ light sources, rather than adjusting the individual chromaticity values of each light source in (\ref{eq19}). The condition to strictly radiate a target white color with chromaticity values $(x_{\mathrm{w}},y_{\mathrm{w}})$ from the $i$-th headlight for all $U$ Rx vehicles is represented as
\begin{align}\label{eq22}
\begin{bmatrix}
( \bt_{X,1}^i - x_{\mathrm{w}} \bt_{T,1}^i )^{\mathrm{T}} \\
\vdots \\
( \bt_{X,U}^i - x_{\mathrm{w}} \bt_{T,U}^i )^{\mathrm{T}} \\
( \bt_{Y,1}^i - y_{\mathrm{w}} \bt_{T,1}^i )^{\mathrm{T}} \\
\vdots \\
( \bt_{Y,U}^i - y_{\mathrm{w}} \bt_{T,U}^i )^{\mathrm{T}}
\end{bmatrix} \bp^i = \bT_{\mathrm{w}}^i\bp^i = \mathbf{0}_{2U},
\end{align}
with the sum RGB ratio constraint $\mathbf{1}_{3}^{\mathrm{T}}\bp_m^i = 1$ for all $m\in \{ 1,\cdots,N_{\mathrm{t}} \}$, which can be concatenated as
\begin{align}\label{eq23}
\begin{bmatrix}
\mathbf{1}_{3}^{\mathrm{T}} & & \\
& \ddots & \\
& & \mathbf{1}_{3}^{\mathrm{T}}
\end{bmatrix} \bp^i = \tilde{\bI}\bp^i = \mathbf{1}_{N_{\mathrm{t}}}.
\end{align}
The RGB ratios to radiate the target white color can be obtained through the following convex optimization formulated by considering (\ref{eq22}) and (\ref{eq23}) as
\begin{align}
&\mathrm{(P2):}\,\, \underset{\bp^i}{\min}\,\,\, \begin{Vmatrix}
\tilde{\bI} \bp^i - \mathbf{1}_{N_{\mathrm{t}}}
\end{Vmatrix} \notag\\
&\qquad\quad\quad\,\, \mathrm{s.t.}\quad \bT_{\mathrm{w}}^i \bp^i = \mathbf{0}_{2U},\tag{2-a}\\
&\qquad\qquad\quad\quad\,\,\,\,\bp^i \ge \mathbf{0}_{3N_{\mathrm{t}}},\tag{2-b}
\end{align}
for $i\in \{ 1,2 \}$. In Section \ref{sec_AO}, we initialize the RGB ratios $\bp^i$ for several target white colors by (P2) to attain the best SLNR-based precoders.

\subsection{Relaxation of white light constraint}\label{relax}
The white light constraint in Section \ref{white_const} may be too strict to set the RGB ratios since we only focus on a point on the CIE 1931 $xy$ coordinates for the stable illumination, resulting in significant power attenuation due to the absorption by GNPs. In this subsection, we propose to relax the white light constraint to improve the performance of SLNR-based precoder. Even though the white light is generally defined as light that appears colorless to the human eye, it can be mapped to various points on the CIE 1931 $xy$ coordinates. This mapping is based on the color temperature, which depends on the color emitted depending on the temperature of black body. The SPD per area of ideal black body with the wavelength $\lambda$ and temperature $K$ is obtained with the Plank's law as
\begin{equation}
S_{\mathrm{BB}}(\lambda,K) = \frac{ 2h_{\mathrm{P}}c_{\mathrm{s}}^2 }{ \lambda^5 ( e^{ \frac{ h_{\mathrm{P}}c_{\mathrm{s}} }{ \lambda k K } } - 1 ) },
\end{equation}
where $h_{\mathrm{P}}$ is the Plank constant, $c_{\mathrm{s}}$ is the speed of light, and $k$ is the Boltzmann constant. The tri-stimulus values by the ideal black body with the temperature $K$ are represented as
\begin{align}
X_{K}&= \int_{\lambda} S_{\mathrm{BB}}(\lambda,K)\bar{x}(\lambda) d\lambda, \notag\\
Y_{K}&= \int_{\lambda} S_{\mathrm{BB}}(\lambda,K)\bar{y}(\lambda) d\lambda, \notag\\
Z_{K}&= \int_{\lambda} S_{\mathrm{BB}}(\lambda,K)\bar{z}(\lambda) d\lambda.
\end{align}
The CIE 1931 $xy$ coordinates $(x_K,y_K)$ for the white color corresponding to a color temperature $K$ can be obtained in the same manner as in (\ref{eq19}). We will consider multiple white light such as cool, neutral, and warm white within a finite white color temperature set $\mathcal{K}$.

\begin{figure}
	\centering
	\includegraphics[width=1.0\columnwidth]{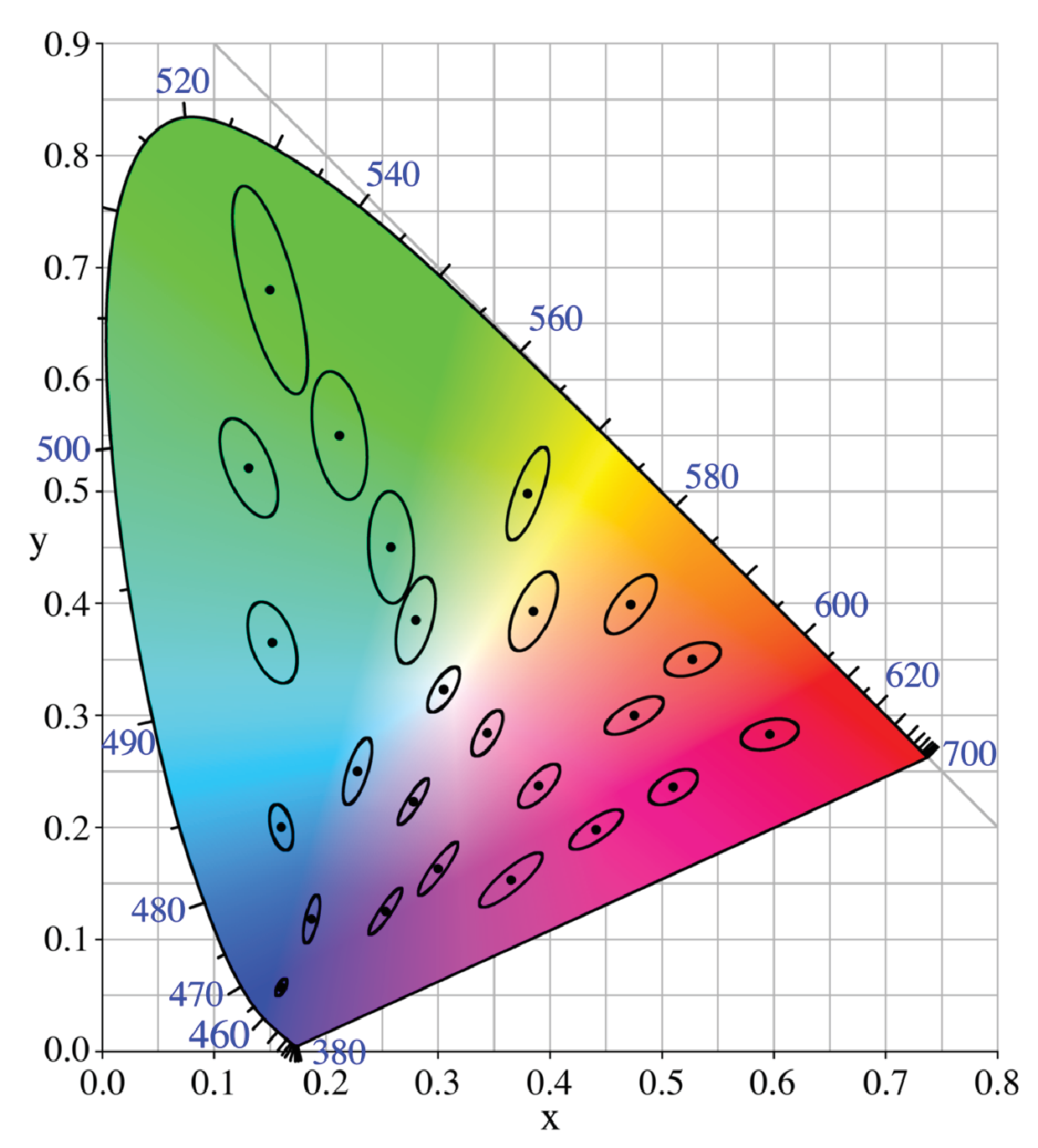}
	\caption{MacAdam ellipses on the CIE 1931 $xy$ coordinates where the figure is obtained from \cite{Jiang:2016}.}\label{MacAdam}
\end{figure}

Various colors around a target color with specific chromaticity values on the CIE 1931 $xy$ coordinates are indistinguishable from the target color to the human eye. The indistinguishable colors can be contained in an ellipse, i.e., the MacAdam ellipse \cite{MacAdam:1942}, with the permissible deviation as depicted in Fig. \ref{MacAdam}, which is the standard deviation of color matching (SDCM) officially used in LED lighting to indicate the light color deviation. The regions by the MacAdam ellipses for multiple temperatures are challenging to be indicated clearly on the CIE 1931 $xy$ coordinates since they are overlapped and irregular. Instead, we employ the quadrangle of the MacAdam ellipse for the color temperature $K$ that can be described with four linear inequalities as \cite{Ansi:2015}
\begin{align}\label{eq26}
\begin{bmatrix}
a_{1,K} & b_{1,K} \\
\vdots & \vdots \\
a_{4,K} & b_{4,K}
\end{bmatrix}\begin{bmatrix}
x_u^i \\
y_u^i
\end{bmatrix} + \begin{bmatrix}
c_{1,K}\\
\vdots\\
c_{4,K}
\end{bmatrix} \le \mathbf{0}_{4},
\end{align}
for each $u\in \{ 1,\cdots,U \}$ and $i\in\{ 1,2 \}$, where $a_{\beta,K}$, $b_{\beta,K}$, and $c_{\beta,K}$ are the coefficients of linear inequalities consisting of a quadrangle for $\beta\in\{ 1,\cdots,4 \}$. Using (\ref{eq20}), the quadrangle region in (\ref{eq26}) is represented as
\begin{align}\label{eq27}
\bT_{u,K}^i\bp^i  \le \mathbf{0}_{4}
\end{align}
with
\begin{align}
\bT_{u,K}^i = \begin{bmatrix}
a_{1,K}(\bt_{X,u}^i)^{\mathrm{T}}+b_{1,K}(\bt_{Y,u}^i)^{\mathrm{T}}+c_{1,K}(\bt_{T,u}^i)^{\mathrm{T}}\\
\vdots\\
a_{4,K}(\bt_{X,u}^i)^{\mathrm{T}}+b_{4,K}(\bt_{Y,u}^i)^{\mathrm{T}}+c_{4,K}(\bt_{T,u}^i)^{\mathrm{T}}\\
\end{bmatrix}.
\end{align}
The relaxed white color constraint in (\ref{eq27}) substitutes for the strict constraint in (\ref{eq22}) to improve the sum rate in the following RGB ratio optimization.

\subsection{RGB ratio optimization}
In this subsection, we optimize the RGB ratios $\bp^i$ to maximize the sum SLNR metric taking the relaxed white light constraint discussed in the previous subsection. To express the SLNR in (P1) as a function of $\bp^i$ with given precoders $\mathbf{f}_u^i=\begin{bmatrix} f_{u,1}^i, \cdots, f_{u,N_{\mathrm{t}}}^i \end{bmatrix}^{\mathrm{T}}$, we have
\begin{align}\label{eq29}
\tilde{\bH}_u^i \mathbf{f}_u^i &= \bH_u^i \diag( \bG_u^i \bp^i ) \mathbf{f}_u^i \notag\\
&= \bH_u^i \begin{bmatrix}
(\bg_{u,1}^i)^{\mathrm{T}}\bp_1^i & & \\
& \ddots & \\
& & (\bg_{u,N_{\mathrm{t}}}^i)^{\mathrm{T}}\bp_{N_{\mathrm{t}}}^i
\end{bmatrix} \mathbf{f}_u^i \notag\\
&= \bH_u^i \begin{bmatrix}
f_{u,1}^i ( \bg_{u,1}^i )^{\mathrm{T}} & & \\
& \ddots & \\
& & f_{u,N_{\mathrm{t}}}^i ( \bg_{u,N_{\mathrm{t}}}^i )^{\mathrm{T}}
\end{bmatrix} \bp^i \notag\\
&= \bH_u^i \bG_u^i \bF_u^i \bp^i\notag\\
&= \bar{\bH}_u^i \bF_u^i \bp^i
\end{align}
with
\begin{align}
\bF_u^i= \begin{bmatrix}
f_{u,1}^i \bI_{3}  & & \\
& \ddots & \\
& & f_{u,N_{\mathrm{t}}}^i \bI_{3}
\end{bmatrix}.
\end{align}
Using (\ref{eq29}), the sum SLNR maximization problem for a color temperature $K$ can be represented as
\begin{align}
&\mathrm{(P3):}\,\,\underset{\bp^i,i\in\{ 1,2 \}}{\max}\,\,\,\sum_{i=1}^{2}\sum_{u=1}^{U}\frac{ \mu_{\mathrm{S},u}^i(\bp^i) }{ \mu_{\mathrm{L},u}^i(\bp^i) + \mu_u(\bp^i) + \sigma_{\mathrm{th}}^2 } \notag\\
&\qquad\qquad\,\,\, \mathrm{s.t.}\quad \tilde{\bI}\bp^i \le \mathbf{1}_{N_{\mathrm{t}}}\quad \forall i,\tag{3-a} \label{const: sum power}\\
&\qquad\qquad\qquad\,\,\,\,\,\bT_{u,K}^i\bp^i \le \mathbf{0}_{4}\quad \forall u,i,\tag{3-b} \label{const: white light}\\
&\qquad\qquad\qquad\,\,\,\,\,\bp^i \ge \mathbf{0}_{3N_{\mathrm{t}}}\quad \forall i, \tag{3-c}\label{const: positive}
\end{align}
where
\begin{align}
\mu_{\mathrm{S},u}^i(\bp^i) &= \alpha^2 P_{\mathrm{Tx}}^2(\bp^i)^{\mathrm{T}} \bF_u^i (\bar{\bH}_u^i)^{\mathrm{T}} \bar{\bH}_u^i \bF_u^i \bp^i,\notag\\
\mu_{\mathrm{L},u}^i(\bp^i) &= \alpha^2 P_{\mathrm{Tx}}^2 (\bp^i)^{\mathrm{T}} \bF_u^i \left( \sum_{v\ne u} (\bar{\bH}_{v}^i)^{\mathrm{T}} \bar{\bH}_v^i \right) \bF_u^i \bp^i,\notag\\
\mu_u( \bp^i ) &= \gamma P_{\mathrm{Tx}} \sqrt{ \| \tilde{\bH}_u^{i}\mathbf{1}_{N_{\mathrm{t}}} \|^2 + \alpha^2 \sum_{v=1}^U \| \tilde{\bH}_u^{i}  \mathbf{f}_v^{i} \|^2 }\notag\\ 
&= \gamma P_{\mathrm{Tx}} \sqrt{ \| \bar{\bH}_u^{i} \bp^{i} \|^2 + \alpha^2 \sum_{v=1}^U \| \bar{\bH}_u^{i} \bF_v^{i} \bp^{i} \|^2 }.
\end{align}
In (P3), we also relax the sum RGB ratio constraint (\ref{eq23}) as \eqref{const: sum power} to optimize $\bp^i$ in a larger feasible set. The basic programming methods in CVX cannot be adopted since the objective function of (P3) is not convex for the RGB ratios $\bp^i$. Instead, we can solve the nonconvex problem (P3) with the quadratic transform for multiple ratio fractional programming in \cite{Shen:2018} and SCA method in \cite{Razaviyayn:2014}. By the quadratic transform, the objective function of (P3) for a specific $u$ and $i$ is equivalent to
\begin{align}\label{eq32}
\xi_u(\bp^i) = 2\nu_u^i\sqrt{ \mu_{\mathrm{S},u}^i(\bp^i) } - (\nu_u^i)^2 \left( \mu_{\mathrm{L},u}^i(\bp^i) + \mu_u(\bp^i) + \sigma_{\mathrm{th}}^2 \right),
\end{align}
where $\nu_u^i$ is an auxiliary variable, which is updated to maximize the surrogate function of (\ref{eq32}). Then, the SCA method is adopted after linearizing $\sqrt{ \mu_{\mathrm{S},u}^i(\bp^i) }$ and $\mu_u(\bp^i)$ to the first-order derivatives, which are defined at the $\ell$-th iteration of SCA method as
\begin{align}\label{eq33}
& \biggl( \sqrt{ \mu_{\mathrm{S},u}^i(\bp^i) } \biggr)_{(\ell)} = \notag\\
&\qquad\qquad\,\, \sqrt{ \mu_{\mathrm{S},u}^i(\bp_{(\ell)}^i) } + \frac{ \alpha^2 P_{\mathrm{Tx}}^2(\bp_{(\ell)}^i)^{\mathrm{T}} \bA_u^i ( \bp^i - \bp_{(\ell)}^i ) }{ \sqrt{ \mu_{\mathrm{S},u}^i(\bp_{(\ell)}^i) } },\notag\\
& \biggl( \mu_u(\bp^i) \biggr)_{(\ell)} = \mu_u(\bp_{(\ell)}^i) + \frac{ \gamma^2 P_{\mathrm{Tx}}^2 ( \bp_{(\ell)}^{i} )^{\mathrm{T}} \bB_u^i ( \bp^i - \bp_{(\ell)}^i ) }{ \mu_u(\bp_{(\ell)}^i) }
\end{align}
with
\begin{align}
\bA_u^i &= \bF_u^i (\bar{\bH}_u^i)^{\mathrm{T}} \bar{\bH}_u^i \bF_u^i,\notag\\
\bB_u^i &= ( \bar{\bH}_u^{i} )^{\mathrm{T}} \bar{\bH}_u^{i} + \alpha^2 \sum_{v=1}^{U} \bF_v^i ( \bar{\bH}_u^i )^{\mathrm{T}} \bar{\bH}_u^i \bF_v^i.
\end{align}
In (\ref{eq33}), $\bp_{(\ell)}^i$ is the local point of the $\ell$-th iteration in the SCA method. The surrogate function of (\ref{eq32}) for the $\ell$-th iteration is given as
\begin{align}
&\xi_{u,(\ell)}(\bp^i) = 2\nu_{u,(\ell)}^i \biggl( \sqrt{ \mu_{\mathrm{S},u}^i(\bp^i) } \biggr)_{(\ell)} \notag\\
&\qquad\qquad\,\, - \left( \nu_{u,(\ell)}^i \right)^2 \biggl( \mu_{\mathrm{L},u}^i(\bp^i) + \biggl( \mu_u(\bp^i) \biggr)_{(\ell)} + \sigma_{\mathrm{th}}^2 \biggr).
\end{align}
The surrogate optimization problem for the $\ell$-th iteration is finally given as
\begin{align}
&\mathrm{(P3.1):}\,\,\underset{\bp^i,i\in\{ 1,2 \}}{\max}\,\,\,\sum_{i=1}^{2}\sum_{u=1}^{U} \xi_{u,(\ell)}(\bp^i) \notag\\
&\qquad \mathrm{s.t.}\quad \nu_{u,(\ell)}^i = \frac{ \sqrt{ \mu_{\mathrm{S},u}^i(\bp_{(\ell)}^i) } }{ \mu_{\mathrm{L},u}^i(\bp_{(\ell)}^i) + \mu_u(\bp_{(\ell)}^i) + \sigma_{\mathrm{th}}^2 }\quad \forall u,i \tag{3.1-a} \label{const: auxiliary ell}, \notag\\
&\qquad\qquad\, \eqref{const: sum power},\eqref{const: white light},\eqref{const: positive}.\notag
\end{align}
The RGB ratio optimization problem in (P3.1) can be simply solved by the quadratic programming (QP).

\begin{algorithm}[t]
	\begin{algorithmic} [1]
		\caption{Sum SLNR maximization by alternating optimization (AO)}
		\State \textbf{Initialization:} Set $\ell_1=0$ and a stopping threshold $\epsilon$.
            \Repeat
                \State Set $K$, $a_{\beta,K}$,$b_{\beta,K}$, and $c_{\beta,K}$ for $\beta\in\{ 1,\cdots,4 \}$.
    		\Repeat
                    \State Set $\ell_1 = \ell_1 + 1$.
                    \If{\ell_1 = 1}
                        \State Initialize $\bp^i=\bp_{\mathrm{init}}^i$ using (P2).
                    \Else
                        \State $\bp^i_{\mathrm{init}} = \bp_{(\ell_2)}^i$ and $\bp^i = \bp^i_{\mathrm{init}}$.
                    \EndIf
                    \State Update $\mathbf{f}_u^i$ as the solution of (P1).
                    \State Update $\xi_u(\bp^i)$ with $\bp^i$ and $\mathbf{f}_u^i$.
                    \State Set $\ell_2=1$.
                    \State Set $\bp_{(\ell_2)}^i=\bp_{\mathrm{init}}^i$.
        		\Repeat
                        \State Set $\ell_2 = \ell_2 + 1$.
                        \State Update $\bp_{(\ell_2)}^i$ as the solution of (P3.1).
        		\Until \ $ \begin{vmatrix} \sum_i \sum_u \xi_{u,(\ell_2)}(\bp^i) - \xi_{u,(\ell_2-1)}(\bp^i) \end{vmatrix} < \epsilon$.
                \Until \ $ \begin{vmatrix} \sum_i \sum_u \xi_{u,(\ell_2)}(\bp^i) - \xi_{u}(\bp^i) \end{vmatrix} < \epsilon $.
            \Until \ calculating the maximized sum SLNR for all color temperatures $K\in\mathcal{K}$.
	\end{algorithmic}
\end{algorithm}

\section{Alternating Optimization}\label{sec_AO}
We need to obtain the SLNR-based precoder considering the spectral transmittance by chiroptical properties of GNPs. The SLNR-based precoders $\mathbf{f}_u^i$ are designed by solving (P1) after initializing the RGB ratios $\bp^i$ using (P2). Then, the RGB ratios are optimized in the quadrangle of color temperature $K$ by solving (P3.1) with the SCA method, where the chromaticity values of optimized RGB ratios $\bp^i$ on the CIE 1931 $xy$ coordinates would gradually move to points with less absorption by GNPs as demonstrated in Section \ref{sec_simul}. These SLNR-based precoder and RGB ratio optimization problems are alternatingly solved for each color temperature $K$. The overall algorithm of the proposed technique is represented in Algorithm 1.

We also analyze the complexity of AO to clearly show the practicality of proposed method. The AO iterates solving the generalized eigenvalue problem in (P1) and the convex QP problem with linear constraints via the SCA approach in (P3.1). The complexities of both problems are obtained as $\mathcal{O}( N_{\mathrm{t}}^3 )$ for the generalized eigenvalue problem \cite{Golub:2013} and $\mathcal{O}( N_{\mathrm{t}}^{3} )$ for the QP problem \cite{Luenberger:1984}. With the stopping threshold $\epsilon$, the overall complexity of AO for the worst case can be computed as $\mathcal{O}( N_{\mathrm{t}}^{3.5} \log( \epsilon^{-1} ) )$, which has the polynomial complexity that is more manageable than the exponential growth as the input size increases.

\begin{figure}
	\centering
	\includegraphics[width=1.0\columnwidth]{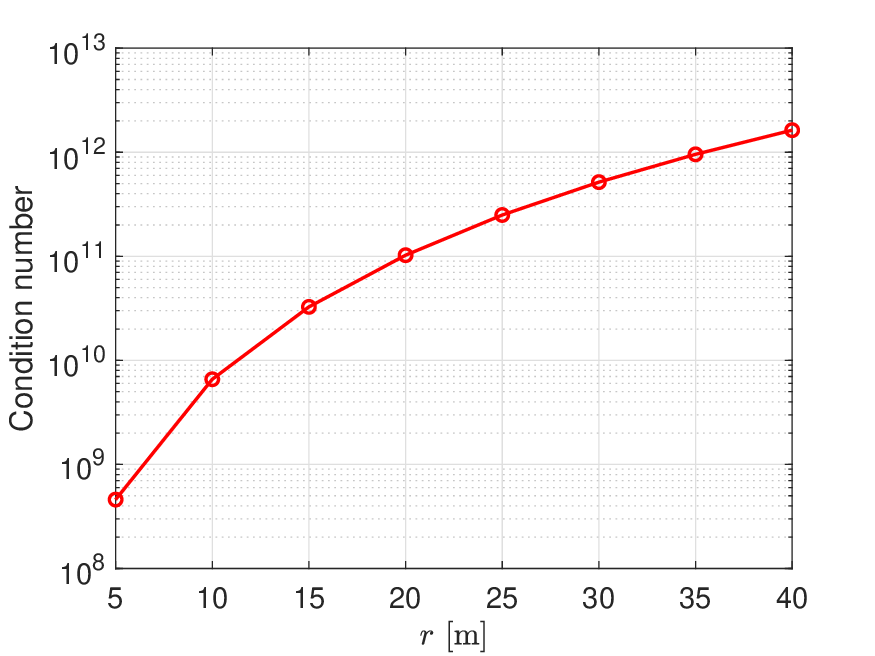}
	\caption{Condition numbers depending on the distance between the headlight of Tx vehicle located in (-1, 0, 1.1) m and the center of PDs at Rx vehicle located in $(0, r, 0.9)$ m.}\label{cond_num}
\end{figure}

\section{Simulation Results}\label{sec_simul}
In this section, we perform simulations to verify the performance of proposed SLNR-based precoder for multi-user MIMO VVLC systems in realistic vehicular environments. The Tx vehicle has two headlights where each headlight consists of $N_{\mathrm{t}}=4$ LEDs. There are $U=3$ Rx vehicles equipped with $N_{\mathrm{r}}=4$ PDs located at the center of each Rx vehicle's rear-end with the height 0.9 m. The central locations of PDs at the Rx vehicles are (-3, 20, 0.9) m, (0, 23, 0.9) m, and (3, 19, 0.9) m, which are denoted as 1, 2, and 3-rd Rx vehicles in order, assuming the road lane width of 3 m. The two headlights of Tx vehicle are located in (-1, 0, 1.1) m and (1, 0, 1.1) m, which are denoted as 1 and 2-nd headlights in order. The LEDs and PDs are positioned with the spacing of $d_{\mathrm{LED}}=4$ cm and $d_{\mathrm{PD}}=1$ cm in a square configuration. We use the spectral responsivity of PD $R_{\mathrm{PD}}(\lambda)$ in \cite{Hranilovic:2006} and the SPDs of light sources $S_c(\lambda)$ with the center wavelengths (450, 521, 630) nm in \cite{Cui:2016}. The spectral transmittance by one cell of GNP plate depending on the azimuth angle of incident light is shown in Fig. \ref{Trans_GNP}. In the simulation results, we adopt several parameters as follows: the current-to-light conversion efficiency $\zeta=0.44$ W/A, the modulation index $\alpha=0.5$, the half power angle of LED $\Phi_{1/2} = 20^{\circ}$, the areas of PD and reflector $A_{\mathrm{PD}}=1$ $\mathrm{cm}^2$ and $A_{\mathrm{ref}}=0.04$ $\mathrm{m}^2$, the thermal noise $\sigma_{\mathrm{th}}^2 = -128.76$ dBm \cite{Komine:2004}, the number of cells in a GNP plate $N=4$, and the system bandwidth $B=100$ MHz. The finite white color temperature set $\mathcal{K}$ has its components as $\{2700,3000,3500,4000,4500,5000,5700,6500 \}$ K.\footnote{The white color temperature set $\mathcal{K}$ is an example to verify the effectiveness of proposed SLNR-based precoding, which needs to be properly limited to comply with the vehicular regulations to avoid the glare for drivers of oncoming vehicles.} We consider the aluminum material for the spectral reflectance of Rx vehicle $\rho(\lambda)$ in (\ref{eq6}) \cite{shanks:2016optics}.

\begin{figure}
	\centering
	\includegraphics[width=1.0\columnwidth]{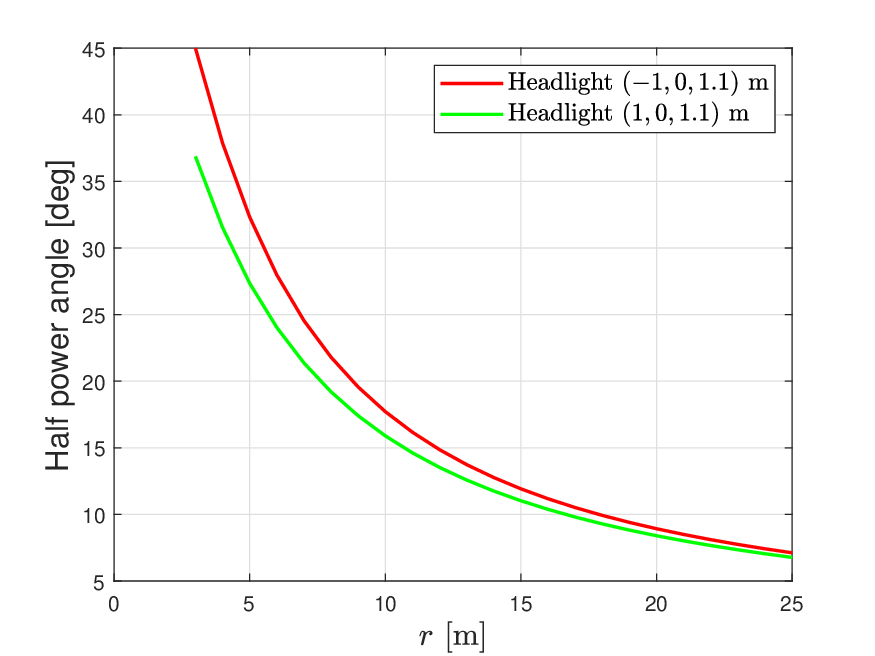}
	\caption{Half power angles of LED $\Phi_{1/2}$ to support Rx vehicles with the locations of $(-3, r+1, 0.9)$ m, $(0, r+4, 0.9)$ m, and $(3, r, 0.9)$ m.}\label{half_p_ang}
\end{figure}

Fig. \ref{cond_num} shows the condition number depending on the distance between a headlight of the Tx vehicle and the center of PDs at an Rx vehicle. The condition number of effective channel is extremely high due to the correlation between PDs, supporting the approximation in (\ref{eq16}). Even though the effective channel has a very high condition number, the LED decorrelation by GNPs makes the nulling of leakage to other Rx vehicles possible, which will be shown shortly. In Fig. \ref{half_p_ang}, we also verify how to set the half power angle of LED $\Phi_{1/2}$ to support multiple Rx vehicles with different distance values $r$. The half power angle of LED $\Phi_{1/2}$ within each headlight should be determined carefully considering the location of Rx vehicle with the largest azimuth angle. For the later simulations with $r=19$ m, the half power angle $\Phi_{1/2}=20^{\circ}$ setting is reasonable since the half power angles of headlights need to be larger than $9.4^{\circ}$ and $8.8^{\circ}$ to support Rx vehicles.

\begin{figure}
	\centering
	\includegraphics[width=1.0\columnwidth]{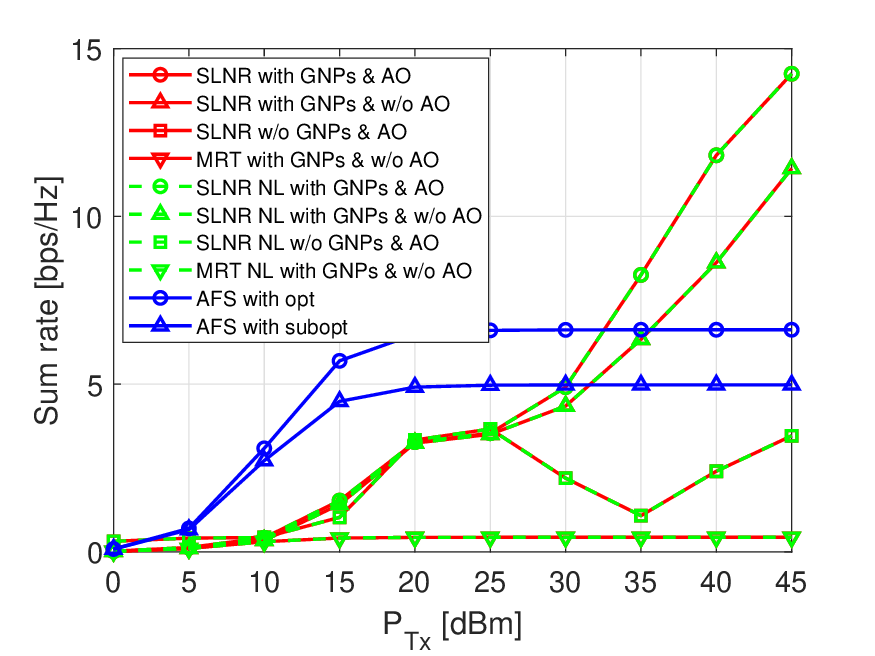}
	\caption{Sum rate performances depending on the Tx power in a multiple access scenario.}\label{MA_sum_Ray_SCA_SLNR}
\end{figure}

We derive the sum rate of $U$ Rx vehicles by using the closed form of upper bounded shot noise and assuming the lower bound of capacity for the VLC channel as \cite{Yin:2017}
\begin{align}
R \approx \frac{1}{2}\sum_{u=1}^U \log_2\left( 1 + \frac{e}{2\pi} \cdot \frac{ \sum_{i=1}^{2} \alpha^2 P_{\mathrm{Tx}}^2 (\mathbf{f}_u^i)^{\mathrm{T}} (\tilde{\bH}_u^i)^{\mathrm{T}}\tilde{\bH}_u^i \mathbf{f}_u^i }{ \mu_{\mathrm{IN},u} } \right)
\end{align}
with
\begin{align}
\mu_{\mathrm{IN},u} = \sum_{i=1}^{2} \left( \alpha^2 P_{\mathrm{Tx}}^2 \sum_{v\ne u}(\mathbf{f}_v^i)^{\mathrm{T}} (\tilde{\bH}_u^i)^{\mathrm{T}}\tilde{\bH}_u^i \mathbf{f}_v^i + \tilde{\sigma}_{\mathrm{sh},u,i}^2 \right) + \sigma_{\mathrm{th}}^2.
\end{align}

In Fig. \ref{MA_sum_Ray_SCA_SLNR}, we first compare the two cases of precoding with different levels of channel knowledge: i) both the LOS and non-LOS paths and ii) only the LOS path. The former case is marked with ``NL'' in the figure. The SLNR-based precoding with the full channel knowledge has negligible gain compared to the case of knowing only the LOS path due to the extremely high path-loss of non-LOS paths, which makes the proposed method more practical. The effect of GNPs is noticeably shown from the sum rate of ``SLNR with GNPs \& w/o AO.'' It outperforms the sum rate of ``SLNR w/o GNPs \& AO,'' which can be considered as a conventional approach that does not mitigate the negative effect of densely positioned LEDs within each headlight, particularly for high SNR regions. GNPs can form different row spaces of effective VVLC channels for multiple users, making it possible to simultaneously maximize signal power and minimize leakage and noise. The proposed method of ``SLNR with GNPs \& AO'' enhances the sum rate more effectively than ``SLNR with GNPs \& w/o AO'' using the RGB ratios for neutral white with the color temperature 4000 K only. Meanwhile, the sum rate of ``SLNR w/o GNPs \& AO'' shows a weird tendency, i.e., it first increases, then decreases, and then increases again until it saturates with $P_{\mathrm{Tx}}$.\footnote{Note that maximizing SLNR does not result in maximizing the sum rate in general.} This is because the channels share almost the same row space that unidirectionally varies the signal and leakage for all users. For mid SNR regions with the Tx power of $[25,35]$ dBm, the sum rate of ``SLNR w/o GNPs \& AO'' decreases since the SLNR-based precoder is not able to simultaneously maximize signal power and minimize leakage due to the overlapped row spaces of the channels. In high SNR regions, ``SLNR w/o GNPs \& AO'' increases the sum rate by using precoders minimizing the leakage power as it becomes dominant. The maximum ratio transmission (MRT) precoder shows limited sum rates even with the decorrelation by GNPs since it solely maximizes the signal power toward each Rx vehicle without accounting for the leakage.

\begin{figure}
	\centering
	\includegraphics[width=1.0\columnwidth]{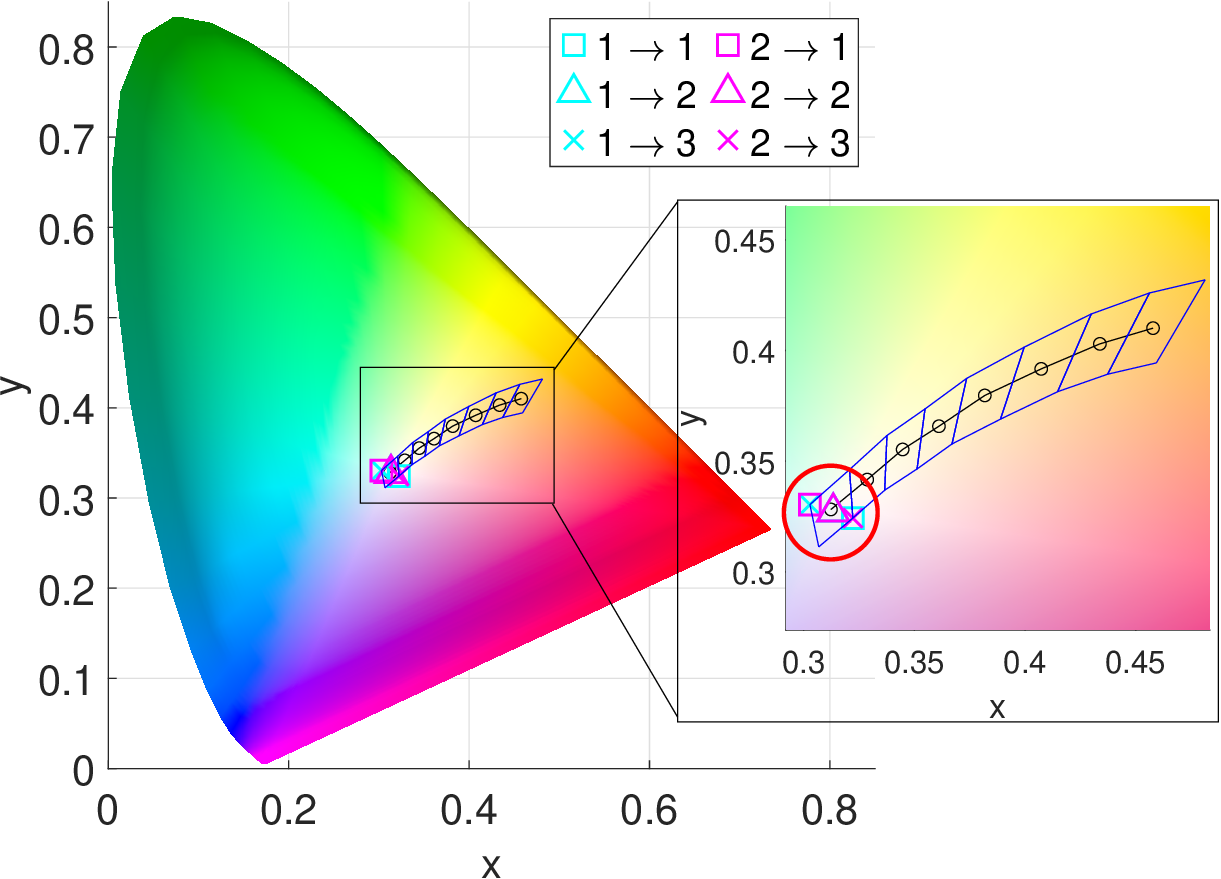}
	\caption{Chromaticity values of optimized RGB ratios on the CIE 1931 $xy$ coordinates in a multiple access scenario.}\label{CIE_Ray_SCA}
\end{figure}

In Fig. \ref{MA_sum_Ray_SCA_SLNR}, we also evaluate the performance of our proposed SLNR-based precoding method in comparison with a state-of-the-art benchmark, i.e., the AFS using matrix headlight. The AFS in \cite{Tebruegge_2:2019} employs matrix headlights with a large number of small LED chips that emit light with the significantly sharp radiation pattern. To support multiple users while mitigating the IUI, the transmitter selects one LED per user that yields the highest received power in a given geometry, i.e., optimal LED. The rest of LEDs do not transmit signals, but they would still consume power for illumination purposes. We assume the same sum power constraint in the proposed VVLC system and the AFS in \cite{Tebruegge_2:2019}, i.e., $N_{\mathrm{t}} P_{\mathrm{Tx}} = N_{\mathrm{t},[21]} P_{\mathrm{Tx},[21]}$ with the number of LEDs used in each matrix headlight $N_{\mathrm{t},[21]}=84$. In Fig. \ref{MA_sum_Ray_SCA_SLNR}, the sum rates of AFS in \cite{Tebruegge_2:2019} show high performances in low SNR regions since the noise is more dominant than the IUI, which is sufficiently suppressed. Even though the IUI is mitigated thanks to the sharp radiation pattern, it becomes non-negligible in high SNR regions by the radiation range of each LED chip itself. The misalignment, i.e., incorrect selection of the optimal LED, may cause noticeable performance degradation. When suboptimal LEDs adjacent to the optimal ones are selected, the misalignment leads to significant performance loss as in ``AFS with subopt.'' This means that the sharp radiation pattern involves a trade-off between the performance gain from the increased SINR and the degradation caused by misalignment. Also, the sum rate would be saturated beyond $P_{\mathrm{Tx},[21]}=30$ dBm, i.e., corresponding to approximately $P_{\mathrm{Tx}}=45$ dBm, due to the intrinsic power limitation of LED chips embedded in matrix headlights. These LED chips can typically support the maximal transmit power of $P_{\mathrm{Tx},[21]}=30$ dBm, constrained by their compact physical size and sophisticated control \cite{Trommer:2019,matrix_HL_1,matrix_HL_2}. In contrast, the proposed method of ``SLNR with GNPs \& AO'' shows lower sum rates due to the absorption by GNPs in low SNR regions. In high SNR regions, however, higher sum rates are achieved since the RGB LEDs for headlights are capable of i) fully using available resources for the IUI mitigation while enhancing the signal power, ii) avoiding potential misalignment issues due to their wide radiation range, and iii) operating at the transmit power of up to 50 W \cite{RGB_LED_1,RGB_LED_2}.

In Fig. \ref{CIE_Ray_SCA}, we analyze the optimized RGB ratios with their chromaticity values on the CIE 1931 $xy$ coordinates. In the figure, the black dots denote the chromaticity values of white light for several color temperatures, and the other indicators, i.e., squares, triangles, and crosses, represent the chromaticity values of optimized RGB ratios for the indistinguishable white light from the $i$-th headlight of Tx vehicle to the $u$-th Rx vehicle denoted as $i\rightarrow u$. To achieve the maximal sum rate gains, the chromaticity values of optimized RGB ratios are positioned in the cool white quadrangle with the color temperature 6500 K due to the reduced absorption by GNPs. The indicators for $1\rightarrow 3$ and $2\rightarrow 1$ are positioned at the edge of quadrangle away from the red color to minimize the absorption by GNPs as much as possible, which is due to the low transmittance by GNPs caused from the large azimuth angle of incident light. The indicators are separated each other to balance the SLNR in the cool white quadrangle.

\begin{figure}
	\centering
	\includegraphics[width=1.0\columnwidth]{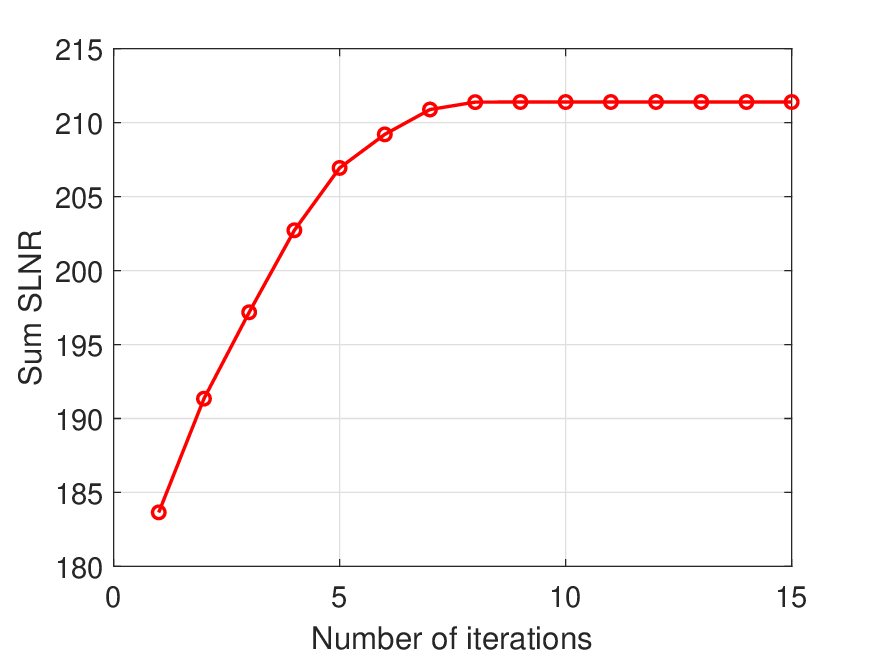}
	\caption{Convergence of AO for the SLNR-based precoding method with $P_{\mathrm{Tx}}=30$ dBm.}\label{conv_AO}
\end{figure}

We analyze the convergence behavior of AO to verify the feasibility of proposed method in Fig. \ref{conv_AO}. Algorithm 1 for AO converges to the optimized sum SLNR within approximately 8 iterations. This ensures that the proposed SLNR-based precoding method computationally practical.

\begin{figure}
	\centering
	\includegraphics[width=1.0\columnwidth]{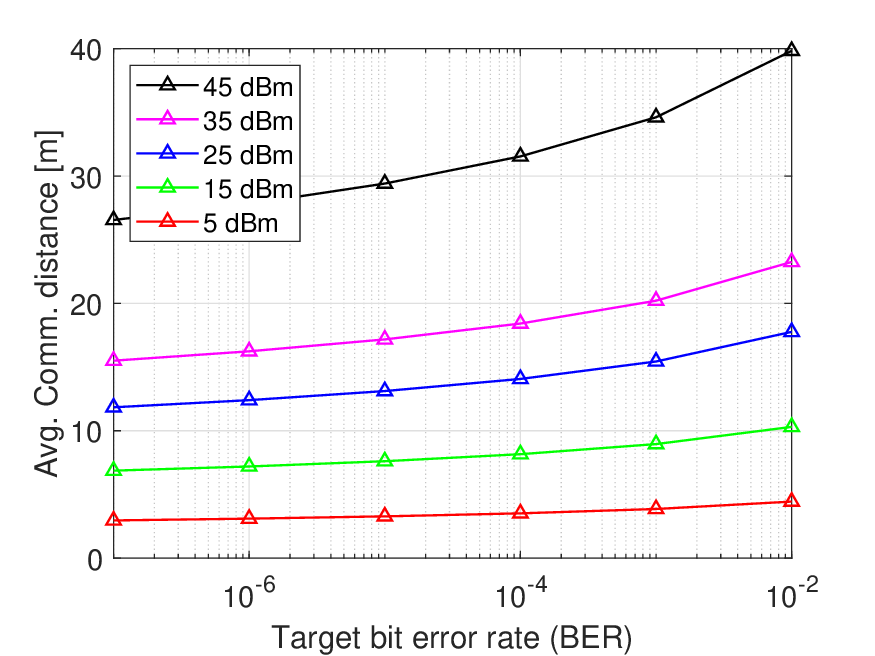}
	\caption{Average communication distance for the target BER.}\label{distance_Ray_SCA}
\end{figure}

We also demonstrate the average communication distance depending on the target BER using the proposed precoder design. The average communication distance is obtained from the distance between the first headlight of the Tx vehicle and the central rear-end of the second Rx vehicle. The effective VVLC channel can be approximated as $\tilde{\bH}_2^1 \approx  (h_2^1)_{\mathrm{L}} \mathbf{1}_{ N_{\mathrm{r}} \times N_{\mathrm{t}} } \diag( \bG_2^1 \bp^1 )$ by negligible non-LOS path-loss and high correlation, where $(h_2^1)_{\mathrm{L}} = \mathbb{E} \{ ( h_{2,nm}^1 )_{\mathrm{L}} \}$ is the LOS path-loss averaged over $n$ and $m$. An upper bound of BER for maximum likelihood (ML) detection with $M$-ary PAM constellation can be obtained as \cite{Alaka:2015}
\begin{align}\label{eq38}
&\mathrm{BER}_M \approx \frac{ 1 }{ M \log_2 M } \sum_{j_1=1}^{M} \sum_{j_2 \ne j_1} d_{\mathrm{H}}( s_{j_1}, s_{j_2} ) \notag\\
&\qquad \times Q\left( \frac{ (h_2^1)_{\mathrm{L}} P_{\mathrm{Tx}} }{ 2 \sigma_2 } \begin{Vmatrix} \mathbf{1}_{ N_{\mathrm{r}}\times N_{\mathrm{t}} } \diag(\bG_2^1 \bp^1) \mathbf{f}_2^1 ( s_{j_2} - s_{j_1} ) \end{Vmatrix} \right),
\end{align}
where $d_{\mathrm{H}}(s_{j_1},s_{j_2})$ is the hamming distance between transmitted symbols $s_{j_1}$ and $s_{j_2}$, and $Q(\cdot)$ is the Q function. The approximated BER in (\ref{eq38}) for $M=4$ is given as
\begin{align}\label{eq39}
\mathrm{BER}_4 \approx \frac{3}{4}Q(\kappa) + Q(2\kappa) + \frac{1}{4}Q(3\kappa)
\end{align}
with
\begin{align}\label{eq40}
\kappa=\frac{ (h_2^1)_{\mathrm{L}} P_{\mathrm{Tx}} d_{4\mathrm{PAM}} }{ 2\sigma_2 } \begin{Vmatrix} \mathbf{1}_{ N_{\mathrm{r}} \times N_{\mathrm{t}} } \diag( \bG_2^1 \bp^1 ) \mathbf{f}_2^1 \end{Vmatrix},
\end{align}
where $d_{4\mathrm{PAM}}$ is the adjacent distance between 4-ary PAM symbols. Using the LOS path-loss of VVLC channel in (\ref{eq5}) and $\kappa$ in (\ref{eq40}), the average communication distance is calculated as
\begin{align}
R_{\mathrm{L}} \approx \biggl( &\frac{ ( n_{\mathrm{LED}} + 1 )A_{\mathrm{PD}} }{ 2\pi }\cos^{n_{\mathrm{LED}}}\phi_{\mathrm{L}}\cos\theta_{\mathrm{L}}\notag\\
&\qquad\,\, \times \frac{ P_{\mathrm{Tx}} d_{4\mathrm{PAM}} }{ 2\sigma_2 \kappa } \begin{Vmatrix}
\mathbf{1}_{ N_{\mathrm{r}} \times N_{\mathrm{t}} } \diag( \bG_2^1 \bp^1 ) \mathbf{f}_2^1
\end{Vmatrix} \biggr)^{\frac{1}{2}}.
\end{align}
Fig. \ref{distance_Ray_SCA} highlights the need for high Tx power to achieve reliable communications using LEDs with the Lambertian radiation pattern. The communication distance will become much longer with the same transmit power if i) the Rx vehicle is equipped with more PDs having larger areas or using single photon avalanche diodes (SPADs) that detect weak signals well \cite{Li:2015}, or ii) the LEDs at the Tx vehicle employ more efficient radiation patterns suitable for vehicular environments \cite{Eldeeb:2021,Karbalayghareh:2020}.

\begin{figure}
	\centering
	\includegraphics[width=1.0\columnwidth]{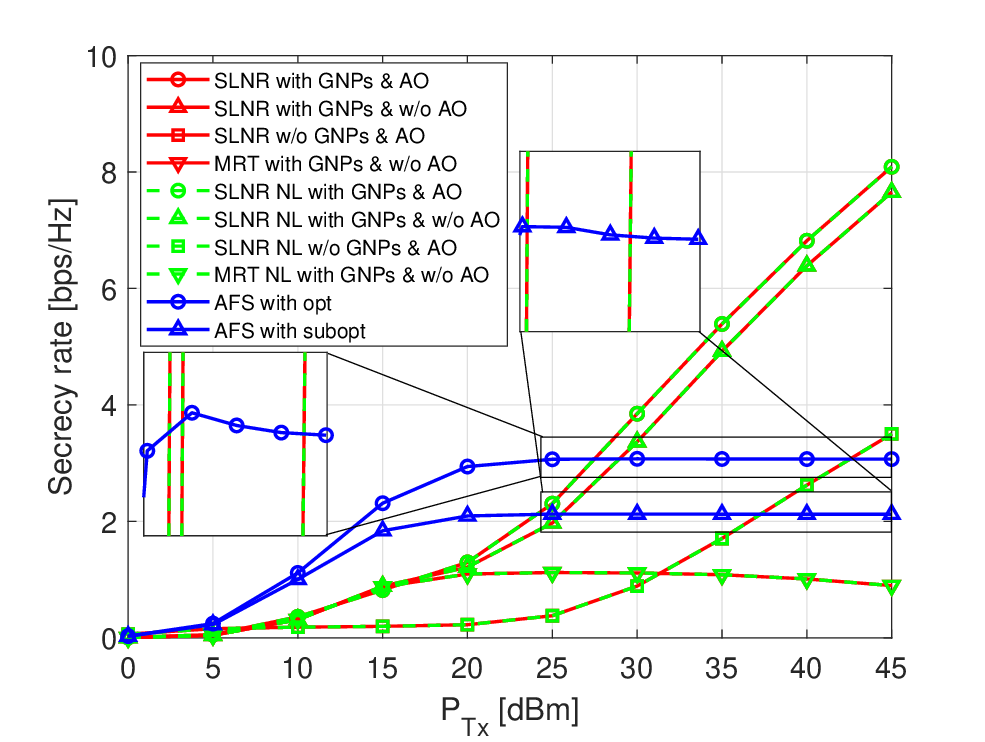}
	\caption{Secrecy rate performances depending on the Tx power in a wiretapping scenario.}\label{sec_Ray_SCA_SLNR}
\end{figure}

\begin{figure}
	\centering
	\includegraphics[width=1.0\columnwidth]{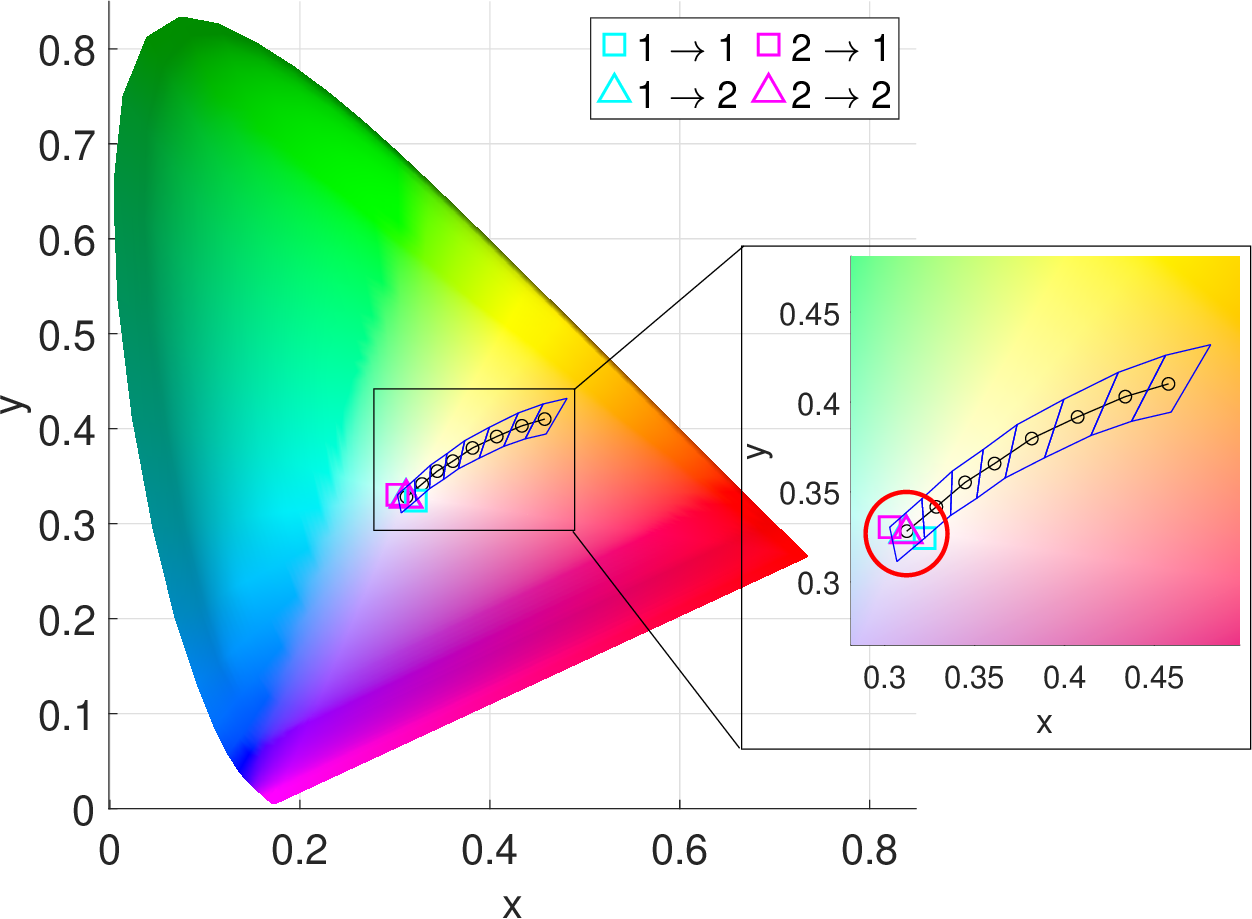}
	\caption{Chromaticity values of optimized RGB ratios on the CIE 1931 $xy$ coordinates in a wiretapping scenario.}\label{CIE_Ray_SCA_sec}
\end{figure}

We finally verify the secrecy rate improvement by proposed SLNR-based precoding method in a wiretapping scenario with a legitimate vehicle (Bob) and an eavesdropping vehicle (Eve). In the wiretapping scenario, the central locations of PDs at Bob and Eve are (-3, 20, 0.9) m and (0, 23, 0.9) m. The other simulation parameters are the same to the multiple access scenario. We assume the Tx vehicle knows the channel state information (CSI) of Eve thanks to estimating geometric information by front sensors. The secrecy rate can be represented as
\begin{align}
R_{\mathrm{s}} &= \max(R_{\mathrm{B}} - R_{\mathrm{E}}, 0),
\end{align}
using the achievable rates of Bob, $R_{\mathrm{B}}$, and Eve, $R_{\mathrm{E}}$, given as
\begin{align}
R_{\mathrm{C}} &\approx \frac{1}{2} \log_2\left( 1 + \frac{e}{2\pi} \cdot \frac{ \sum_{i=1}^{2} \alpha^2 P_{\mathrm{Tx}}^2 (\mathbf{f}^i)^{\mathrm{T}} (\tilde{\bH}_{\mathrm{C}}^i)^{\mathrm{T}}\tilde{\bH}_{\mathrm{C}}^i \mathbf{f}^i }{ \sum_{i=1}^{2} \tilde{\sigma}_{\mathrm{sh,C},i}^2 + \sigma_{\mathrm{th}}^2 } \right)
\end{align}
with
\begin{align}
\tilde{\sigma}_{\mathrm{sh,C},i}^2 = \gamma P_{\mathrm{Tx}} \sqrt{ \| \tilde{\bH}_{\mathrm{C}}^i\bold{1}_{N_{\mathrm{t}}} \|^2 + \alpha^2 \| \tilde{\bH}_{\mathrm{C}}^i \mathbf{f}^i \|^2 },
\end{align}
where $\mathrm{C}\in\{\mathrm{B,E}\}$, $\tilde{\bH}_{\mathrm{C}}^i$ is the effective VVLC channel for $\mathrm{C}$, and $\mathbf{f}^i$ is the proposed SLNR precoder for Bob. In Fig. \ref{sec_Ray_SCA_SLNR}, the secrecy rates when including the non-LOS paths show almost the identical performances to the case where only the LOS path is considered as in Fig. \ref{MA_sum_Ray_SCA_SLNR}. The SLNR-based precoders with GNPs show high secrecy rates as increasing Tx power since they can reduce the leakage toward Eve. ``SLNR w/o GNPs \& AO'' shows different tendency compared to Fig. \ref{MA_sum_Ray_SCA_SLNR}  since the proposed VVLC system for the wiretapping scenario only needs to design a precoder $\mathbf{f}^i$ for each headlight by considering Bob and Eve. The row spaces of VVLC channels are restricted by two users in this scenario, which ensures a larger feasible set different from the multiple access scenario with three users. Despite this, it has a worse secrecy rate compared to other SLNR-based methods for most of SNR regions since Eve can receive intended signals comparable to those at Bob. The RGB ratio optimization further improves the secrecy rate performance as in ``SLNR with GNPs \& AO'' by relatively increasing the signal power more at Bob compared to Eve thanks to the RGB ratios less absorbed by GNPs while reducing the leakage toward Eve. Fig. \ref{CIE_Ray_SCA_sec} shows almost the same chromaticity values of optimized RGB ratios in Fig. \ref{CIE_Ray_SCA} due to the white light constraint that should be satisfied for all front regions of the Tx vehicle while maximizing the sum SLNR. The performance of MRT precoder in Fig. \ref{sec_Ray_SCA_SLNR} demonstrates that the multi-user communications can be significantly vulnerable to eavesdropping as leaked out transmitted signals particularly in high SNR regions. In Fig. \ref{sec_Ray_SCA_SLNR}, the secrecy rates of AFS in \cite{Tebruegge_2:2019} have similar tendencies as in Fig. \ref{MA_sum_Ray_SCA_SLNR}, but it decreases in high SNR regions by eavesdropping. It can be observed in ``AFS with subopt'' that the misalignment further increases the risk of eavesdropping.

\section{Conclusion}\label{sec_conc}
In this paper, we proposed an SLNR-based precoding method with GNPs in vehicular environments to improve the sum rate in a multiple access scenario and the secrecy rate in a wiretapping scenario. GNPs, recently synthesized metamaterials, have chiroptical properties as differential absorption and phase delay for left/right circularly polarized light, which can be controlled by polarization, wavelength, and azimuth angle of incident light. The transmittance according to differential absorption effect of GNPs and cell arrangement with different types, patterns, and sizes of GNPs in a GNP plate was exploited to decorrelate closely positioned LEDs in a headlight of the Tx vehicle. A multi-user MIMO VVLC system with GNPs and RGB light sources within an LED was modelled incorporating the received power representation of shot noise. The SLNR-based precoders were obtained from the generalized Rayleigh quotient with the approximation of shot noise. We also derived the white light constraint for multiple LEDs with the RGB light sources using the generalized color mapping method, and relaxed it to ensure a larger feasible set to optimize the RGB ratios. Then, we optimized the RGB ratios to further enhance the performance gain by reducing the absorption by GNPs under the relaxed white light constraint. The precoders and the RGB ratios were alternatingly updated to maximize the sum SLNR. Finally, we demonstrated through simulation results that the decorrelation effect by GNPs is essential to support multiple Rx vehicles. The proposed SLNR-based precoding would be suitable for vehicular scenarios that demand high data rates and secure communications such as platooning.

As several future works, the adaptive radiation pattern design based on simulation studies needs to be adopted for ensuring the sufficient communication distance in practical VVLC systems, which also can capture the impact of outdoor weather condition and variations across different vehicle types \cite{Eldeeb:2021,Karbalayghareh:2020}. Tail-to-head VVLC using typically lower power compared to head-to-tail VVLC requires further studies to make bidirectional vehicular communications possible. To achieve higher SNR at Rx vehicles in the tail-to-head VVLC, we can consider several approaches, e.g., i) using SPADs and spatial filters tailored to the spectra of signals radiated from taillights, ii) deploying numerous PDs with the large areas to receive signals better, and iii) implementing better combining methods using PDs to improve the SINR. With these methods, the proposed SLNR-based precoding also can be exploited even for multi-user tail-to-head VVLC under a red light constraint considering the taillights emitting the light of red wavelength for warning purposes.

\bibliographystyle{IEEEtran}
\bibliography{refs_all}

\begin{IEEEbiography}[{\includegraphics[width=1in,height=1.25in,clip,keepaspectratio]{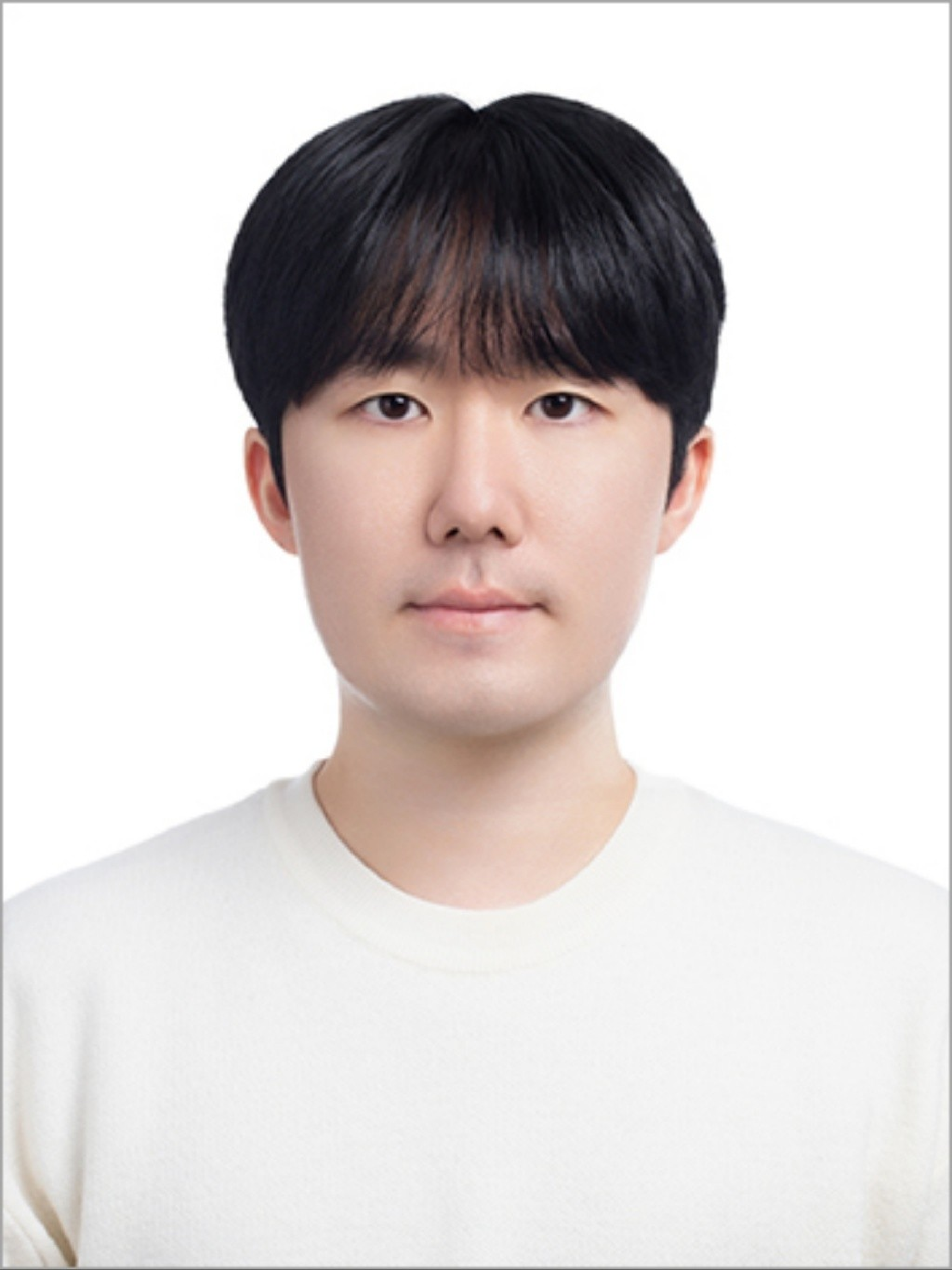}}]{Geonho Han } (Member, IEEE) received the B.S. degree in electrical engineering from the Pohang University of Science and Technology, Pohang, South Korea, in 2018. He earned his Ph.D. degree (M.S./Ph.D. combined course) from the School of Electrical Engineering at KAIST, South Korea, in 2025. He is currently a researcher in Electronics and Telecommunications Research Institute (ETRI).

His research interests include the design and analysis of integrated sensing and communication (ISAC) systems, visible light communication (VLC) systems, vehicular-to-everything (V2X) communications, and extreme massive MIMO (E-MIMO) systems. He was awarded the KAIST EE Best Ph.D. Dissertation Award in 2025.
\end{IEEEbiography}

\begin{IEEEbiography}[{\includegraphics[width=1in,height=1.25in,clip,keepaspectratio]{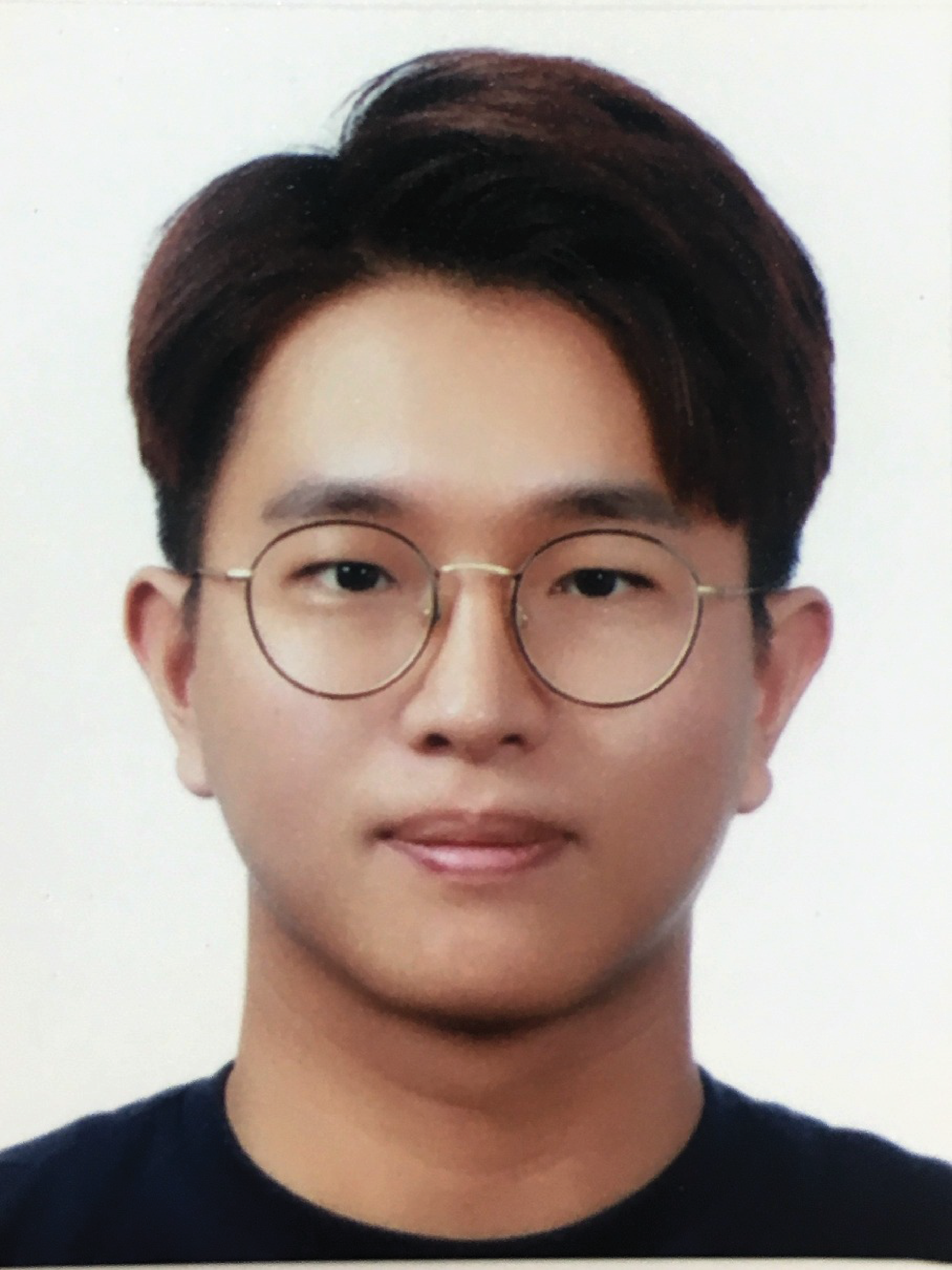}}]{Hyuckjin Choi } (Member, IEEE) received the B.S. and M.S. degrees in electrical engineering from POSTECH, Pohang, South Korea, in 2018 and 2020, respectively, and the Ph.D. degree in electrical engineering from KAIST, Daejeon, South Korea, in 2024.

He is currently a Post-Doctoral Researcher with the Electrical Engineering and Computer Science (EECS) Department, University of California at Irvine. In 2024, he was a Post-Doctoral Researcher with KAIST. He was a recipient of 2023 KAIST EE Best Research Achievement Award (Kim Choong-Ki Award) and NRF Korea Sejong Fellowship from 2024 to 2025. He developed the 3D ray-tracing simulator, WiThRay, for the development and evaluation of wireless communication systems. His current research interests include the design and analysis of reconfigurable intelligent surfaces-aided communications, visible light communications, and communication systems utilizing machine-learning techniques.
\end{IEEEbiography}

\begin{IEEEbiography}[{\includegraphics[width=1in,height=1.25in,clip,keepaspectratio]{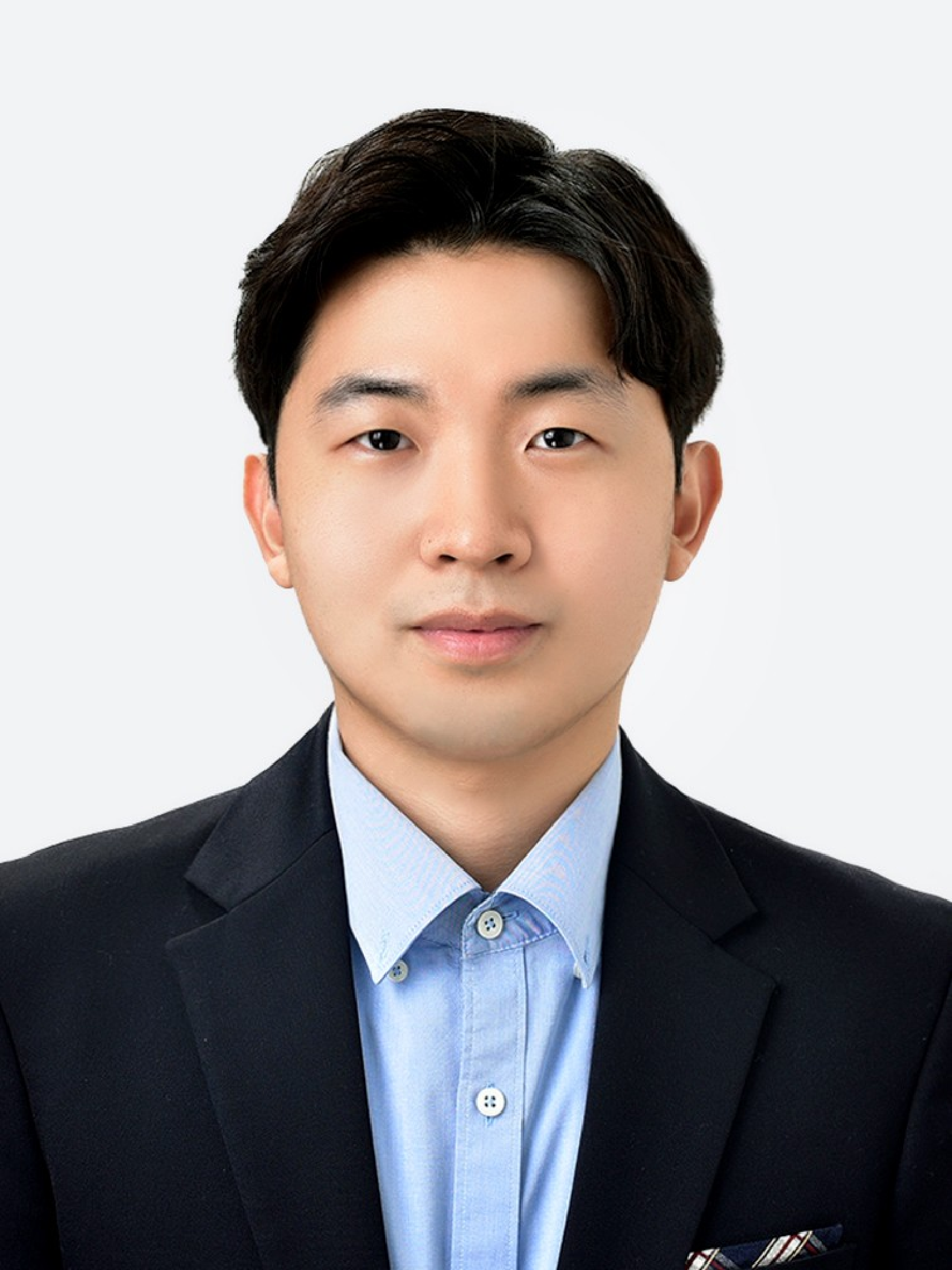}}]{Hyesang Cho } (Member, IEEE) received the B.S., M.S., and Ph D. degrees in electrical engineering from Korea Advanced Institute of Science and Technology (KAIST) in 2019, 2021, and 2024. He is currently a post-doctoral researcher with KAIST. His research interests are in the design and analysis of reconfigurable intelligent surface, rate-splitting multiple access, mmWave/THz communication systems, and unmanned aerial vehicles. He was a recipient of the 14th Electronic News ICT Outstanding Paper Award and the Best Paper Award of the Korean Institute of Communications and Information Sciences (KICS) Summer Conference, 2020. He was a recipient of the 2025 KAIST Jang Young Sil Fellow program (Jang Young Sil Postdoctoral Researcher Track).
\end{IEEEbiography}

\begin{IEEEbiography}[{\includegraphics[width=1in,height=1.25in,clip,keepaspectratio]{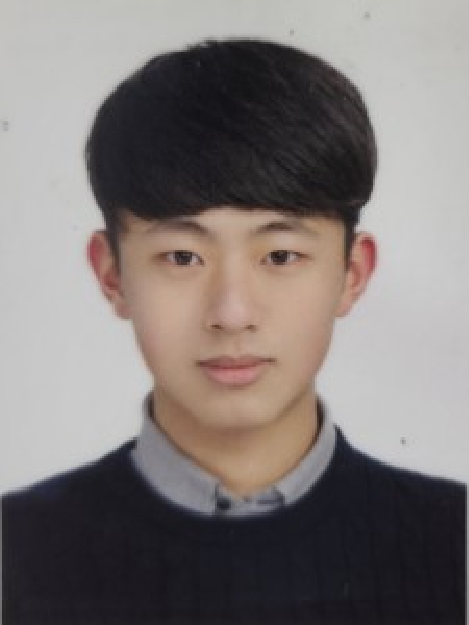}}]{Jeong Hyeon Han } is a Ph.D. candidate in materials science and engineering at Seoul National University under the supervision of Prof. Ki Tae Nam. He received his B.S. from the same institution in 2021. His research focuses on the design, manipulation, and interpretation of chiral light–matter interactions in plasmonic systems.
\end{IEEEbiography}

\begin{IEEEbiography}[{\includegraphics[width=1in,height=1.25in,clip,keepaspectratio]{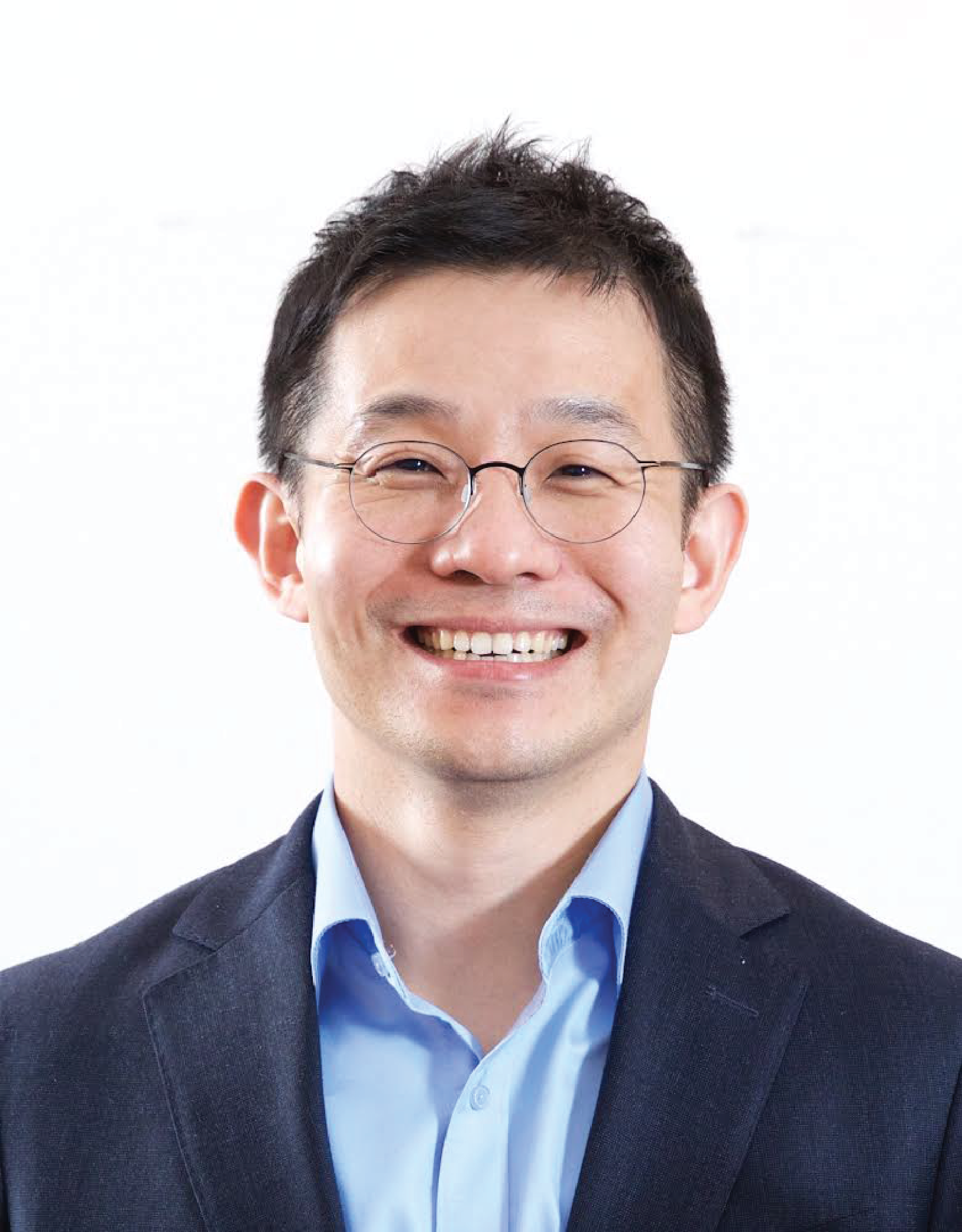}}]{Ki Tae Nam } has been a professor at Seoul National University since 2010. He received his Ph.D. from Massachusetts Institute of Technology in 2007 and carried out postdoctoral research at Lawrence Berkeley National Laboratory. His research focuses on bioinspired strategies for the design and synthesis of functional materials, particularly for energy conversion and optical applications.

His group pioneered the peptide-based synthesis of chiral gold nanoparticles, establishing a new class of inorganic chiral nanostructures. Their work revealed how molecular-level asymmetry can be encoded into plasmonic materials, as demonstrated in their publications in Nature (2018) and Nature (2022).
\end{IEEEbiography}

\begin{IEEEbiography}[{\includegraphics[width=1in,height=1.25in,clip,keepaspectratio]{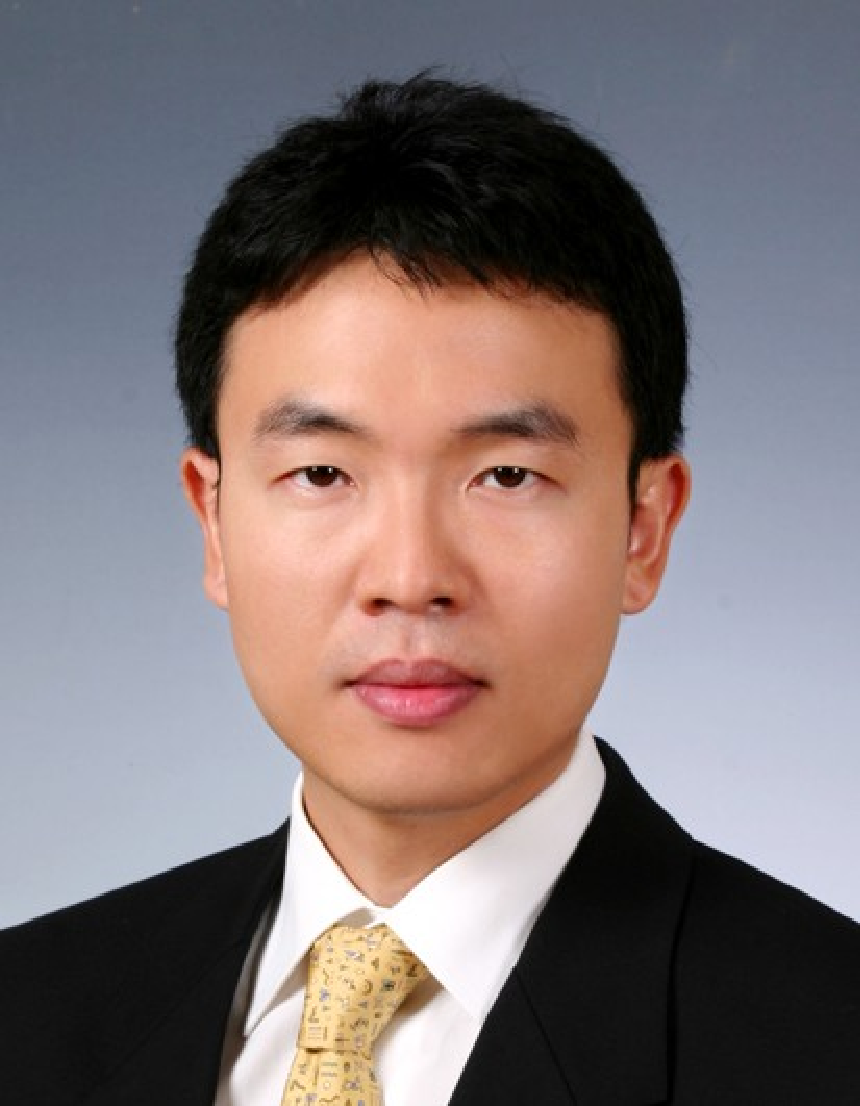}}]{Junil Choi } (Senior Member, IEEE) received the B.S. (Hons.) and M.S. degrees in electrical engineering from Seoul National University in 2005 and 2007, respectively, and the Ph.D. degree in electrical and computer engineering from Purdue University in 2015.

He is currently working as an (Named) Ewon Associate Professor with the School of Electrical Engineering, KAIST. From 2007 to 2011, he was a member of Technical Staff at the Samsung Advanced Institute of Technology (SAIT) and Samsung Electronics Company Ltd., South Korea, where he contributed to advanced codebook and feedback framework designs for the 3GPP LTE/LTE-Advanced and IEEE 802.16m standards. Before joining KAIST, he was a Post-Doctoral Fellow at The University of Texas at Austin from 2015 to 2016 and an Assistant Professor at POSTECH from 2016 to 2019. His research interests include the design and analysis of massive MIMO, mmWave communications, satellite communications, visible light communications, and communication systems using machine-learning techniques.

Dr. Choi was a co-recipient of the 2022 IEEE Vehicular Technology Society Best Vehicular Electronics Paper Award, the 2021 IEEE Vehicular Technology Society Neal Shepherd Memorial Best Propagation Award, the 2019 IEEE Communications Society Stephen O. Rice Prize, the 2015 IEEE Signal Processing Society Best Paper Award, and the 2013 Global Communications Conference (GLOBECOM) Signal Processing for Communications Symposium Best Paper Award. He was awarded the Michael and Katherine Birck Fellowship from Purdue University in 2011, the Korean Government Scholarship Program for Study Overseas (2011–2013), the Purdue University ECE Graduate Student Association (GSA) Outstanding Graduate Student Award in 2013, the Purdue College of Engineering Outstanding Student Research Award in 2014, the IEEE ComSoc AP Region Outstanding Young Researcher Award in 2017, the NSF Korea and Elsevier Young Researcher Award in 2018, the KICS Haedong Young Researcher Award in 2019, and the IEEE Communications Society Communication Theory Technical Committee Early Achievement Award in 2021. He is an IEEE Vehicular Technology Society Distinguished Lecturer, the Area Editor of IEEE Open Journal of the Communications Society Signal Processing for Communications and an Associate Editor of IEEE Transactions on Wireless Communications and IEEE Transactions on Communications.
\end{IEEEbiography}

\end{document}